\documentclass{elsart3p}

\usepackage{graphics}
\usepackage{graphicx}
\usepackage{epsfig}

\usepackage{amsmath,amssymb}

\usepackage{units}
\usepackage{multicol}
\usepackage{multirow}
\usepackage{subfigure}
\usepackage{textcomp}
\usepackage{xspace}

\makeatletter
\def\cleareads{%
  \csname email@text\endcsname={}%
  \csname url@text\endcsname={}%
  \csname corauth@text\endcsname={}%
  \global\chardef\has@ead@email=0\relax
  \global\chardef\has@ead@url=0\relax
  }
\makeatother

\def\Cerenkov{Cherenkov\xspace}
\def\eref#1{(\Ref{#1})\xspace}
\def\Eref#1{Equation\,(\Ref{#1})\xspace}
\def\Erefs#1{Equations\,(\Ref{#1})\xspace}
\def\sref#1{Sect.\,\Ref{#1}\xspace}
\def\srefs#1{Sects.\,\Ref{#1}\xspace}
\def\fref#1{Fig.\,\Ref{#1}\xspace}

\def\tref#1{Tab.\,\Ref{#1}\xspace}
\def\gcm2{g/cm$^2$\xspace}

\def\deg{$^\circ$\xspace}

\newtoks\pnam
\def\Label#1{\label{\the\pnam#1}}
\def\Ref#1{\ref{\the\pnam#1}}
\def\Cite#1{\cite{\the\pnam#1}}
\def\Bibitem#1{\bibitem{\the\pnam#1}}

\def\etal{{\it et al.}\xspace}
\def\theseproc#1{Proc. of the 5th Fluorescence Workshop, Nucl. Instr. \&
                 Meth. A (2008) in press.}

\begin{document}

\begin{frontmatter}

% Title, authors and addresses

% use the thanksref command within \title, \author or \address for footnotes;
% use the corauthref command within \author for corresponding author footnotes;
% use the ead command for the email address,
% and the form \ead[url] for the home page:
% \title{Title\thanksref{label1}}
% \thanks[label1]{}
% \author{Name\corauthref{cor1}\thanksref{label2}}
% \ead{email address}
% \ead[url]{home page}
% \thanks[label2]{}
% \corauth[cor1]{}
% \address{Address\thanksref{label3}}
% \thanks[label3]{}

\title{Air Fluorescence Relevant for Cosmic-Ray Detection ---
       Summary of the 5th Fluorescence Workshop, El Escorial 2007}

% use optional labels to link authors explicitly to addresses:
% \author[label1,label2]{}
% \address[label1]{}
% \address[label2]{}

\author[m]{Fernando Arqueros},
\author[n]{J\"org R. H\"orandel},
\author[k]{Bianca Keilhauer}

\address[m]{Facultad de Ciencias F\'{\i}sicas,
Universidad Complutense de Madrid, E-28040 Madrid, Spain}
\address[n]{Radboud Universiteit Nijmegen, Department of Astrophysics,
             P.O. Box 9010, 6500 GL Nijmegen, The Netherlands}
\address[k]{Universit\"at Karlsruhe, Institut f\"ur Experimentelle
            Kernphysik, Postfach 3640, 76021 Karlsruhe, Germany}

\begin{abstract}
High-energy cosmic rays with energies exceeding $10^{17}$~eV are
frequently observed by measurements of the fluorescence light induced
by air showers.  A major contribution to the systematic uncertainties
of the absolute energy scale of such experiments is the insufficient
knowledge of the fluorescence light yield of electrons in air.  The
aim of the 5th Fluorescence Workshop was to bring together
experimental and theoretical expertise to discuss the latest progress
on the investigations of the fluorescence light yield.  The results of
the workshop will be reviewed as well as the present status of
knowledge in this field.  Emphasis is given to the fluorescence light
yield important for air shower observations and its dependence on
atmospheric parameters, like pressure, temperature, and humidity. The
effects of the latest results on the light observed from air showers
will be discussed.
\end{abstract}

%\journal{Nuclear Instruments and Methods A}
\end{frontmatter}

\section{Introduction}
\Label{intro}

The Earth is permanently exposed to a vast flux of particles from
outer space. Most of these particles are fully ionized atomic nuclei,
covering a large range in energy from the MeV regime to energies above
$10^{20}$~eV.  In 1962 the first event with an energy exceeding
$10^{20}$~eV was recorded \Cite{linsley}.  Such cosmic rays are the
highest-energy particles in the Universe, carrying the (macroscopic)
energy of about 50~J concentrated on a single nucleus. Since their
first discovery more than 40 years ago their origin has been an open
question. ``How do cosmic accelerators work and what are they
accelerating?'' is one of eleven science questions for the new century
asked by the National Research Council of the National Academies of
the United States \Cite{quarkscosmos}, underlining the importance of
this topic to astroparticle physics.

The properties of cosmic rays at highest energies are investigated
with various experiments
\Cite{naganowatson}\Cite{bergmanbelz}\Cite{kamperttaup07}\Cite{jrhwuerzburg}.
The flux of cosmic rays with energies exceeding $10^{20}$~eV is below
1 particle per square kilometer and century. Thus, a reasonable
measurement of these particles requires huge detectors operated stably
over long periods of time.  At present, this can be realized only with
ground based experiments, registering the secondary particles
generated by high-energy cosmic rays in the atmosphere.  When a
high-energy cosmic ray enters the Earth's atmosphere it induces a
cascade of secondary particles, an extensive air shower.  By far the
most abundant particles in air showers are photons, electrons, and
positrons, comprising the electromagnetic shower component.  On their
way through the atmosphere the (relativistic) charged particles emit
\Cerenkov radiation and excite nitrogen molecules to emit fluorescence
light. A small fraction of the secondary particles eventually reaches
the observation level.  Over time, several methods to measure
extensive air showers have been established. They can be divided into
two groups: experiments measuring secondary particles (electrons,
muons, and hadrons) reaching ground level and detectors observing the
emitted \Cerenkov or fluorescence photons.  The latter allow for a
three-dimensional reconstruction of the shower profile in the
atmosphere.

A critical issue for all experiments is to establish an absolute
energy scale for the measured showers. For experiments registering
secondary particles at ground level this usually involves the usage of
simulations of the shower development in the atmosphere, thus,
introducing systematic uncertainties due to our limited knowledge of
the hadronic interactions at such high energies.

On the other hand, fluorescence measurements of air showers provide a
calorimetric measurement of the energy deposited in the air, being
(nearly) independent of air shower simulations.  The deposited energy
is assumed to be proportional to the energy of the primary, shower
inducing, particle. At present, this is the most direct and model
independent method to determine the energy of an air shower. The main
systematic uncertainty of this method arises from the insufficient
knowledge of the fluorescence light yield of electrons in air.
However, this is a quantity which can be measured in laboratory
experiments, injecting electrons into air targets.  Responsible for
the fluorescence emission of the nitrogen molecules is mainly the
electromagnetic shower component. The critical energy of electrons in
air is about 84~MeV, thus, the bulk of particles has energies easily
accessible at accelerators, or, at lower energies, even through
radioactive $\beta$-decays.

This is the main focus of the present article. It gives an overview on
the actual status of the know\-ledge of the fluorescence light yield
of electrons in air, important for air shower detection.  It
summarizes the results of the 5th Fluorescence Workshop, which
was held in El Escorial, Spain from September 16th to 20th,
2007.  After a short overview on the principle of the detection of air
showers using fluorescence light (\sref{sec:principle}), experiments
applying this technique to register cosmic-ray induced air showers are
described (\sref{sec:eashistory}).  The main physical processes
involved in the production of fluorescence light in air are reviewed
in \sref{sec:fl_physics}.  Contemporary experimental tools and
theoretical treatments are discussed in \sref{sec:status}.  Latest
data are compiled in \sref{sec:data}, paying special attention to the
dependence of the fluorescence light yield on atmospheric parameters,
such as pressure, temperature, and humidity.  In the following section
(\sref{sec:altitude}) the influence of the results obtained on the
light yield in air showers developing in realistic atmospheres is
discussed. An outlook describing the next steps in determining the
fluorescence yield in air concludes the article (\sref{sec:outlook}).
The most important pioneering measurements of the fluorescence light
yield are summarized in an accompanying article \Cite{histref}.

\section{The fluorescence technique}
\Label{sec:fl_technique}

\subsection{Principle of air shower detection with fluorescence light}
\Label{sec:principle}

A high-energy cosmic ray entering the atmosphere induces a cascade of
secondary particles. One way to determine the energy of the primary
particle is to measure the energy deposited in an  absorber (i.e.\ the
atmosphere), this is called a calorimetric energy measurement (e.g.\
\Cite{fabjan}). If the shower is absorbed completely, the energy of
the primary particle is identical to the energy deposited.  However,
in an air shower some energy is escaping a calorimetric measurement: a
fraction of secondary particles reaches ground level and some energy
is carried away by ``invisible" particles such as neutrinos.  Luckily,
corrections for this effect are small and rather model independent, as
will be discussed below.  The calorimetric measurement by means of
fluorescence light detection uses the fact that secondary particles in
showers (mostly electrons and positrons) deposit energy in the
atmosphere by ionization or excitation of air molecules. The excited
nitrogen molecules subsequently relax to their ground state partially
by the emission of fluorescence photons. The light is emitted
isotropically, which implies that showers can be viewed from the side,
thus, telescopes can observe large fiducial volumes of air.  Most of
the fluorescence light is emitted in the near UV region with
wavelengths between about 300 and 400~nm.  Simulation studies show
that most of the energy deposited into the atmosphere arises from
electrons (and positrons) with energies below 1~GeV with a maximum at
about 30~MeV \Cite{risseheck}. It is commonly assumed that the number
of emitted fluorescence photons is proportional to the energy
deposited in the atmosphere.

The number of fluorescence photons $\d N_\gamma$ which are generated
in a layer of atmosphere with thickness $\d X$ registered by a
fluorescence detector can be expressed as
\footnote{In contrast to later on, we use in this section the fluorescence
          yield per unit wavelength interval.}
\begin{equation}\Label{ngammaeas}
 \frac{\d N_\gamma}{\d X}=\int \frac{\d^2 N_\gamma^0}{\d X \,\d \lambda} \cdot
   \tau_{atm}(\lambda,X) \cdot \varepsilon_{FD}(\lambda) \d\lambda .
\end{equation}
$\varepsilon_{FD}$ denotes the efficiency of the fluorescence detector
and $\tau_{atm}$ the transmission of the atmosphere. The latter
includes all transmission losses due to optical absorption, Rayleigh
scattering, and Mie scattering from the point of emission to the
detector.  The number of emitted fluorescence photons $\d N_\gamma^0$
emitted per wavelength interval $\d\lambda$ and matter traversed $\d
X$ is obtained as
\begin{eqnarray}
 \frac{\d^2 N_\gamma^0}{\d X\,\d\lambda}=&\int Y(\lambda,P,T,u,E)
   \cdot \frac{\d N_e(X)}{\d E}
   \\ \nonumber & ~~~~~~~~~~~~~~~~~~~~~
   \cdot \frac{\d E_{dep}}{\d X} \d E .
\end{eqnarray}

The energy spectrum of the electrons (and positrons) at an atmospheric
depth $X$ is given by $\d N_e(X)/\d E$ and $\d E_{dep}/\d X$ describes
the energy deposited in a layer of atmosphere  with thickness $\d X$.
The fluorescence light yield $Y$ describes the number of emitted
photons per deposited energy (photons per MeV). For a calorimetric
measurement we are interested in the deposited energy, thus, this
definition relates the searched quantity directly to the observed
amount of light. In the literature $Y$ is frequently given in units of
photons per meter (or photons per unit length). This definition has
the disadvantage that the number of photons emitted per unit length
changes with varying air density. It is non-trivial to convert the two
quantities into each other. Throughout this article we use the
definition of photons per deposited energy, unless noted otherwise.

The fluorescence light yield $Y$ at a wavelength $\lambda$ depends on
the atmospheric pressure $P$, the temperature $T$, the humidity $u$,
and, in principle, as well on the energy of the electrons $E$.  If the
light yield is assumed to be energy independent, it can be taken out
of the integral yielding
\begin{equation}
 \frac{\d^2 N_\gamma^0}{\d X\,\d\lambda}= Y(\lambda,P,T,u)
  \cdot\frac{\d E_{dep}^{tot}}{\d X} .
\end{equation}
The total energy deposited in an atmospheric layer with thickness $\d X$ is
written as
\begin{equation}
  \frac{\d E_{dep}^{tot}}{\d X} =\int \frac{\d N_e(X)}{\d E}
   \cdot \frac{\d E_{dep}}{\d X} \d E .
\end{equation}
With \eref{ngammaeas} the relation
\begin{eqnarray}
 \frac{\d N_\gamma}{\d X}=\frac{d E_{dep}^{tot}}{\d X}
     \int Y(\lambda,P,T,u) \cdot \tau_{atm}(\lambda,X) \\
     \cdot \varepsilon_{FD}(\lambda) \d\lambda \nonumber
\end{eqnarray}
is obtained. It shows that the number of fluorescence photons detected
is proportional to the energy deposited in the atmosphere.  It remains
to be shown that the fluorescence yield is indeed independent of the
electron energy, see \sref{sec:Edep} below.

To calculate the energy of the primary particle from the observed
fluorescence light still some corrections are necessary.  The observed
light contains also a contamination of \Cerenkov light, either direct
light (mostly emitted in forward direction) or scattered light.  This
effect has to be corrected for on an event-to-event basis
\Cite{nerling}.  It has to be taken into account that the cascade is
not absorbed completely in the atmosphere and secondary particles
reach ground level (longitudinal leakage of the calorimeter).
Furthermore, particles which are not detected (neutrinos, high-energy
muons) carry away energy. This ``invisible" energy depends slightly on
the mass of the primary particle and the hadronic interaction model
used to describe the shower development in the atmosphere.
Investigations of the Auger Collaboration indicate that the correction
factor varies between 1.07 and 1.17 only, assuming primary protons or
iron nuclei and applying different hadronic interaction models
\Cite{pierog}.

For illustration, we discuss here the systematic uncertainties of the
absolute energy scale of the Pierre Auger Observatory (as evaluated
before the 5th Fluorescence Workshop) \Cite{augersyst}.  The
fluorescence telescopes are end-to-end calibrated using a large
homogeneous light source which leaves an uncertainty of 9.5\%.
Uncertainties in the shower reconstruction contribute with 10\% and
the correction of the invisible energy adds another 4\% to the error
budget. The atmospheric profile above the observatory is regularly
monitored \Cite{keilhauer2004}. However, an uncertainty of about 4\%
for the energy scale of an individual event remains.  In addition,
uncertainties related to the fluorescence yield have to be taken into
account. The dominant contribution is due to the absolute light yield
(14\%). The dependence on atmospheric parameters contributes with 7\%.
This results in a total systematic uncertainty of 22\%.  Values for
other experiments are similar and confirm that the biggest uncertainty
for the absolute energy scale is the insufficient knowledge of the
fluorescence yield $Y(\lambda,p,T,u)$. To review the latest progress
on the determination of this value is the objective of the present
article.

\subsection{Air shower experiments applying the fluorescence technique}
\Label{sec:eashistory}

In the following, the principle set-ups of air shower detectors
applying the fluorescence technique are briefly sketched.
Illustratively, a fluorescence detector for such an experiment has to
be able to observe a 100~W light bulb\footnote{The fluorescence light
of a $10^{17}$~eV shower corresponds to a light bulb of about 100~W.}
moving at the speed of light through the atmosphere watched
from a distance of 30~km.  To realize this, large-aperture telescopes
are used to focus the light on cameras equipped with fast
photomultiplier tubes, sensitive in the near UV region.
\footnote{Recent results of air shower experiments applying the
fluorescence technique are summarized in e.g.\
\Cite{naganowatson}\Cite{bergmanbelz}\Cite{kamperttaup07}\Cite{jrhwuerzburg}.}
First ideas to use the Earth's atmosphere as vast scintillation
detector were discussed in the early 1960s \Cite{suga}.  The early
history of experiments applying the fluorescence technique is
summarized elsewhere \Cite{tanahashi}.

\subsubsection{Initial experiments}

The pilot experiment to study the feasibility of detecting air showers
with the fluorescence technique was the ``Cornell Wide Angle System"
proposed and built by K.~Greisen and colleagues in the 1960s
\Cite{bunner1967}. It consisted of three detector stations set up in
the vicinity of the Cornell University campus.  Each station comprised
five photomultiplier tubes in a hexa\-hedron arrangement, with a tube
pointing north, south, east, west, and upward, respectively.  The four
radial tubes were tilted upwards by 30\deg.  The system was
operational for about 1000 hours. Light flashes were recorded, but
they could not be attributed to air showers beyond doubt.

In 1967, a full scale fluorescence experiment was constructed by
Greisen's group. It comprised 500 photomultiplier tubes, each
corresponding to a pixel with a solid angle of 0.01 sr. The
photomultipliers were divided into ten modules, each of them was
equipped with a 0.1 m$^2$ Fresnel lens. The experiment was operated
for several years but was not sensitive enough to detect high-energy
cosmic rays.

Similar activities were conducted by the Tokyo group leading to the INS-Tokyo
experiment. In 1969 they recorded first clear fluorescence light signals from
an extensive air shower with an energy exceeding $5\cdot10^{18}$~eV
\Cite{hara}.

\subsubsection{The Fly's Eye experiment}

In 1976 physicists from the University of Utah detected fluorescence
light from cosmic-ray air showers. Three prototype modules were used
at the site of the Volcano Ranch air shower array near Albuquerque,
New Mexico. Each module comprised a 1.8~m diameter mirror for light
collection with a camera consisting of 14 photomultiplier tubes at the
focal plane.  Fluorescence light was recorded in coincidence with
an air shower array. These prototypes led eventually to the
development of the Fly's Eye detector.

The Fly's Eye observatory \Cite{flyseye} consisted of two stations,
separated by 3.3~km.  The first one (Fly's Eye~1) comprised 67 front
aluminized spherical section mirrors, with a diameter of 157~cm.
Winston light collectors and photomultipliers were hexagonally packed
in groups of either 12 or 14 light sensing units, or ``eyes" mounted
in the focal plane of each mirror.  The photomultipliers (EMI 9861B)
had a fairly uniform quantum efficiency over the spectral range from
310 to 440~nm.  A motorized shutter system kept the ``eyes" both light
tight and weather proof during the day and permitted exposure to the
sky at night. Each mirror unit was housed in a single, motorized
corrugated steel pipe about 2.13~m long and 2.44~m in diameter. The
units were turned down with mirror and open end facing the ground
during the day and turned up at night to a predetermined position so
that each ``eye" observed an angular region of the sky. In total, 880
``eyes" were observing the complete upper hemisphere.  The projection
of each hexagonal ``eye" onto the celestial sphere resembles the
compound ``eye" of an insect, hence, the name Fly's Eye.  The second
telescope (Fly's Eye~II) was a smaller array of identical units, with
120 ``eyes" in total, observing roughly one azimuthal quadrant of the
night sky with elevation angles ranging between 2\deg and 38\deg above
the horizon.  Whenever the first telescope recorded an event, it sent
an infrared flash of light towards the second telescope, which
recorded pulse integrals and arrival times. The shower track geometry
was reconstructed either from hit patterns and timing information by a
single Fly's Eye detector or by stereoscopic viewing and relative
timing by both Fly's Eyes.  The experiment has been operated between
1981 and 1993. For the first time the fluorescence technique has been
applied successfully to explore the properties of ultra high-energy
cosmic rays on a large scale.

\subsubsection{The HiRes experiment}

The High Resolution Fly's Eye experiment (HiRes) was located in Utah,
USA (40$^\circ$ N, 112$^\circ$ W) \Cite{hiresexp}. It was the successor
of the Fly's Eye experiment.  HiRes consisted of two detector sites
(Hires I \& II) separated by 12.6~km, providing almost 360\deg
azimuthal coverage, each.  Both telescopes were formed by an array of
detector units. The mirrors consisted of four segments and formed a
5.1~m$^2$ spherical mirror. At its focal plane an array of
$16\times16$ photomultiplier tubes was situated, viewing a solid angle
of $16^\circ\times16^\circ$.  HiRes~I consisted of 22 detectors,
arranged in a single ring, overlooking between 3\deg and 17\deg in
elevation. This detector used an integrating ADC read-out system,
which recorded the photomultiplier tubes' pulse height and time
information.  HiRes~II comprised 42 detectors, set up in two rings,
looking between 3\deg and 31\deg in elevation. It was equipped with a
10~MHz flash ADC system, recording pulse height and timing information
from its phototubes.  The experiment has been operated between 1997
and 2007.

\subsubsection{Telescope Array}

The Telescope Array is an air shower experiment in the West Desert of
Utah (USA), 140 miles south of Salt Lake City (39.3\deg N, 112.9\deg
W) \Cite{telescopearray}.  It comprises 576 scintillator stations and
three fluorescence detector sites on a triangle with about 35~km
separation.  Each fluorescence detector station is equipped with 12 to
14 telescopes, viewing $3^\circ-33^\circ$ in elevation and
$\approx108^\circ$ in azimuth \Cite{telescopearrayfd}.  The telescopes
have a segmented spherical mirror with a diameter of 3.3~m and a focal
length of 3.0~m. Each telescope has a camera comprising 256
photomultiplier tubes (Hamamatsu R9508), corresponding to a pixel size
of about 1\deg each. The sensitive area of a camera is
$1~\mbox{m}\times1~\mbox{m}$, which corresponds to a field of view of
15\deg in elevation and 18\deg in azimuth. The photomultipliers are
read out by a FADC system. The experiment has been taking data since 2007.

\subsubsection{The Pierre Auger Observatory}

The observatory combines the observation of fluorescence light with
imaging telescopes and the measurement of particles reaching ground
level in a ``hybrid approach" \Cite{augerexp}.  The southern site
(near Malarg\"ue, Argentina, 35.2\deg S, 69.5\deg W, 1400~m above sea
level) of the worlds largest air shower detector consists of 1600
water \Cerenkov detectors set up in an area covering 3000~km$^2$. Four
telescope systems overlook the surface array.  A single telescope
system comprises six telescopes, overlooking separate volumes of air.
Each telescope is situated in a bay, protected by a remotely operated
shutter.  Light enters the bay through an UV transmitting filter and a
ring of corrector lenses.  A circular diaphragm (2.2~m diameter),
positioned at the center of curvature of a spherical mirror, defines
the aperture of the Schmidt optical system.  A
$3.5~\mbox{m}\times3.5~\mbox{m}$ spherical mirror focuses the light
onto a camera with an array of $22\times20$ hexagonal pixels.  Each
pixel is a photomultiplier tube, complemented by light collectors.
Each camera pixel has a field of view of approximately 1.5\deg in
diameter. A camera overlooks a total field of view of 30\deg azimuth
$\times$ 28.6\deg elevation. The photomultiplier signals are read out
by FADC systems.  The observatory has been completed in 2008 and has
been taking data stably with a growing number of detector stations
since 2004.

Presently (2008), the Pierre Auger Collaboration is extending the
observatory to lower energies. For this objective additional
high-elevation telescopes (HEAT, High Elevation Auger Telescopes) are
being built, covering an angular range from 30\deg to 60\deg in
elevation \Cite{klagesmerida}.  These telescopes will be operated with
an additional infill array of surface detectors combined with
underground muon counters (AMIGA, Auger Muons and Infill for the
Ground Array)~\Cite{AMIGAmerida} and antennae to detect radio
emission from air showers \Cite{radiomerida}.  To complete and extend
the investigations begun in the South, the Pierre Auger Collaboration
presently prepares an observatory in the northern hemisphere in
Colorado, USA \Cite{anorthmerida}. The set-up will, similarly to the
southern site, comprise water \Cerenkov detectors and fluorescence
telescope systems.

\subsubsection{ASHRA}

The All-sky Survey High Resolution Air-shower telescope (ASHRA) is a
proposed detector system to simultaneously measure \Cerenkov and
fluorescence light on the entire sky with 1 arc min resolution
\Cite{ashra}.  It is planned to install two stations at a distance of
about 30 to 40~km on an island in Hawaii. A station comprises 12
wide-angle telescopes. Each telescope has a field of view of
$50^\circ\times50^\circ$, read out by CMOS sensor arrays.

\subsubsection{JEM-EUSO}

JEM-EUSO is a proposed super-wide field UV telescope to detect ultra
high-energy cosmic rays with energies above 10$^{20}$~eV
\Cite{jemeuso}.  It will be attached to the International Space
Station (ISS) and will observe fluorescence photons emitted by air
showers from an orbit of about 430~km altitude.  The three dimensional
development of the shower is reconstructed from a series of images of
the shower. The spatial resolution is about $0.75\times0.75$~km$^2$.
A double Fresnel lens module with 2.5~m diameter is the baseline
optics for the JEM-EUSO telescope. The focal surface is equipped with
about 6000 multi-anode photomultipliers. The launch is planned for
2012.

\section{Physical processes involved in the generation of air-fluorescence
light}
\Label{sec:fl_physics}

Electrons passing through the atmosphere deposit energy due to
inelastic collisions with air molecules. A small fraction of them give
rise to the production of the fluorescence light observed in the
spectral range of interest ($290 - 430$~nm). This air-fluorescence
light is produced by nitrogen molecules.

In this section the main features of the excitation and de-excitation
of N$_2$ molecules will be reviewed (\sref{sec:exc_de-exc}).
Fluorescence quenching including the humidity effect will also be
discussed (\sref{sec:quenching}). Finally, the definition of the
various parameters associated with the fluorescence yield as used in
the literature will be presented (\sref{sec:yield}).

\subsection{Electron excitation and radiative de-excitation}
\Label{sec:exc_de-exc}

\begin{figure}[t]
\centering
 \epsfig{width=0.49\textwidth, file=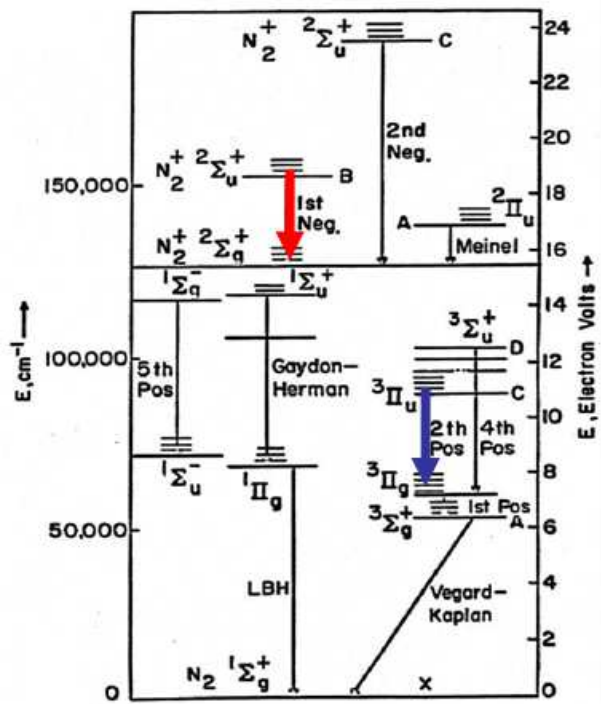}
  \caption{Molecular levels of N$_2$ and N$_2^+$. Broad arrows represent the
           main transitions (1N and 2P systems)~\Cite{bunner1967}.}
\Label{mol-levels}
\end{figure}

A scheme of the molecular levels of N$_2$ and N$_2^+$ is shown in
\fref{mol-levels}. As is well known from elementary molecular physics,
each electronic state is split in vibrational levels $v$. In addition,
each vibrational level is split in rotational sub-levels following a
complicated structure. Electron collision excites molecular nitrogen
in the ground state to upper levels. Down going arrows in
\fref{mol-levels} represent the de-excitation processes giving rise to
fluorescence radiation. Although transitions take place between
individual rotational levels of the upper and lower states, the
corresponding rotational structure of the molecular spectrum is not
resolved in our experiments. Under moderate spectral resolution,
transitions between vibrational levels give rise to molecular bands
$v-v'$ with a spectral width and shape determined by the rotational
structure\footnote{See e.g.\ \Cite{airfly_P} for some illustrative
examples.}. The set of bands connecting a given pair of electronic
states is named a band system.

In our spectral range, nitrogen fluorescence comes basically from the
Second Positive system C$^3\Pi_u$ $\rightarrow$ B$^3\Pi_g$ of N$_2$
and the First Negative system B$^2\Sigma^+_u$ $\rightarrow$
X$^2\Sigma^+_g$ of N$_2^+$ (see \fref{mol-levels}) which in the
air-fluorescence community are usually denoted as 2P and 1N systems,
respectively. Notice that while the 2P system is generated after the
N$_2$ X$^1\Sigma_g^+$ $\rightarrow$ C$^3\Pi_u$ excitation, 1N
fluorescence takes place as a consequence of the X$^1\Sigma_g^+$
$\rightarrow$ (N$_2^+$)B$^2\Sigma^+_u$ molecular ionization, leaving
the nitrogen ion in a specific excited state. The wavelengths of the
molecular bands of nitrogen are well known (see for instance
\Cite{wav_table}).

Apart from the 1N and 2P systems the weak bands of the N$_2$
Gaydon-Herman system (GH) have been observed in the air-fluorescence
spectrum \Cite{GH_system}\Cite{airfly_P}.

\begin{figure}[t] \centering
\epsfig{width=\columnwidth, file=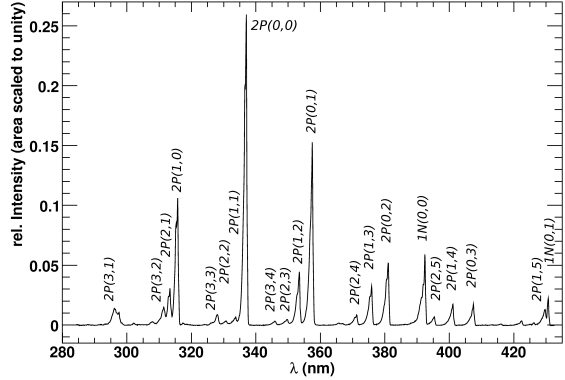}
  \caption{Air fluorescence spectrum excited by 3 MeV electrons at 800 hPa as
           measured by the AIRFLY Collaboration \Cite{5thFW_airfly_P}.}
\Label{airflyspec}
\end{figure}

\begin{table}[t]
 \caption{Transitions and corresponding wavelengths of the air-fluorescence
          spectrum \Cite{5thFW_airfly_P}.~\Label{tab:wavelengths}}
\begin{minipage}{\linewidth}
\begin{center}
  \begin{tabular}{ccccc} \hline
 transition ~~~& $\lambda$(nm) &  $\quad$ & transition ~~~& $\lambda$(nm) \\
 \cline{1-2} \cline{4-5}
2P(3,1) & 296.2 & ~~~~~~~~ & GH(0,5) & 366.1 \\
2P(2,0) & 297.7 &          & 2P(3,5) & 367.2 \\
GH(6,2) & 302.0 &          & 2P(2,4) & 371.1 \\
GH(5,2) & 308.0 &          & 2P(1,3) & 375.6 \\
2P(3,2) & 311.7 &          & 2P(0,2) & 380.5 \\
2P(2,1) & 313.6 &          & 2P(4,7) & 385.8 \\
2P(1,0) & 315.9 &          & GH(0,6) & 387.7 \\
GH(6,3) & 317.6 &          & 1N(1,1) & 388.5 \\
2P(4,4) & 326.8 &          & 1N(0,0) & 391.4 \\
2P(3,3) & 328.5 &          & 2P(2,5) & 394.3 \\
2P(2,2) & 330.9 &          & 2P(1,4) & 399.8 \\
2P(1,1) & 333.9 &          & 2P(0,3) & 405.0 \\
2P(0,0) & 337.1 &          & 2P(3,7) & 414.1 \\
GH(0,4) & 346.3 &          & 2P(2,6) & 420.0 \\
2P(2,3) & 350.0 &          & 1N(1,2) & 423.6 \\
2P(1,2) & 353.7 &          & 2P(1,5) & 427.0 \\
2P(0,1) & 357.7 &          & 1N(0,1) & 427.8 \\ \hline
\end{tabular}
\end{center}
\end{minipage}
\end{table}

A spectrum typically observed at high pressure between 280 and 430~nm
for air is depicted in \fref{airflyspec} \Cite{5thFW_airfly_P}.  The
labels mark 21 major transitions. All important transitions and the
corresponding wavelengths between 290 and 430~nm are compiled in
\tref{tab:wavelengths}.

\begin{figure}[t]
\centering
\epsfig{width=0.49\textwidth, file=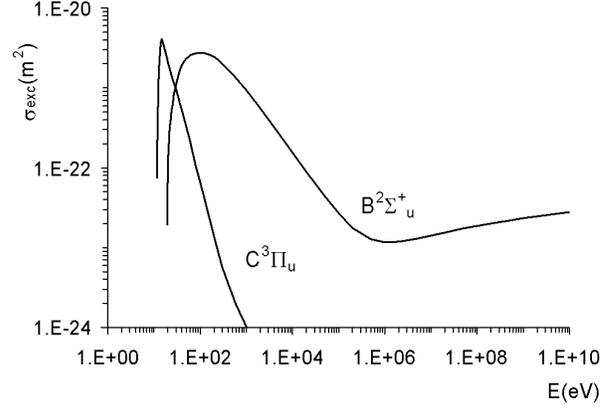}
  \caption{Total cross sections for the excitation of the electronic states
           C$^3\Pi_u$ and B$^2\Sigma^+_u$ versus electron energy 
           \Cite{arqueros_Astr_Ph}.}
\Label{exc_cros_sect}
\end{figure}

The cross section for excitation of the upper electronic levels of
both systems as a function of electron energy is displayed in
\fref{exc_cros_sect}. The curve for the 2P system shows a sharp maximum
at about 15 eV followed by a fast $E^{-2}$ decrease, as expected from
the optically forbidden nature of this transition.  On the contrary,
the excitation cross section for the 1N system shows a much softer
maximum at about 100 eV followed by a much slower $(\log{E})/E$
decrease which becomes a soft growing behavior at relativistic
energies \Cite{blanco}\Cite{arqueros_Astr_Ph}.

For a given electronic state the cross section for the excitation to a
vibrational level $v$ is proportional to the Franck-Condon factor
$q_{X\rightarrow v}$, defined as the overlapping integrals between the
vibrational wave functions of the lower and upper levels of the
excitation process. The Einstein coefficients $A_{vv'}$ give the
probability per unit time of radiative de-excitation $v-v'$.
Therefore, the probability of emission of a fluorescence $v-v'$ photon
by electron impact is proportional to the optical cross section
defined as
\begin{equation}
 \Label{eq:opt_cr_sect}
%%F
\sigma_{vv'} = \sigma_v\frac{A_{vv'}}{\sum_{v'}{A_{vv'}}}=\sigma_v B^{vv'}\,,
% \sigma_{vv'} = \sigma_v\frac{A_{vv'}}{\sum_{v'}{A_{vv'}}}\,,
%%F
\end{equation}
and therefore, in the absence of other effects, the relative intensity
of a molecular band with respect to a reference transition (e.g.\ 0-0)
of the same system is given by
\begin{equation}
 \Label{eq:Rel_I_0}
 \frac{I^0_{vv'}}{I^0_{00}}=\frac{\sigma_{vv'}}{\sigma_{00}} =
 \frac{q_{X\rightarrow v}}{q_{X\rightarrow 0}}\frac{B^{vv'}}{B^{00}}\,.
\end{equation}

Tabulated values for both parameters $q_{X\rightarrow v}$ and
$A_{vv'}$ are available in the literature \Cite{gilmore}\Cite{laux}. The
relative intensities between bands of different systems can also be
predicted using the relative values of the corresponding excitation
cross sections.  Transition probabilities determine the radiative
lifetime $\tau^r$ of the excited level
\begin{equation}
\Label{eq:tau_r} \frac{1}{\tau^r_v} = A_v = \sum_{v'}{A_{vv'}}\,.
\end{equation}
Notice that, as shown below, in a laboratory experiment both relative
intensities \eref{eq:Rel_I_0} and lifetime \eref{eq:tau_r} have to be
corrected by the effect of collisional quenching.

\subsection{Fluorescence quenching}
\Label{sec:quenching}

At high pressure, molecular de-excitation by collision with other
molecules of the medium plays an important role (collisional
quenching). At a given temperature, the corresponding transition
probability $A_c^v$ is proportional to the collision frequency and,
thus, to the gas pressure $P$. The characteristic pressure $P'_v$ is
defined as the one for which the probability of collisional quenching
equals that of radiative de-excitation $A^c_v(P'_v)=A_v$,
\begin{equation}
\Label{eq:A_c} A^c_v (P) = A_v \frac{P}{P'_v}\,.
\end{equation}
Thus, the fluorescence intensity in the absence of quenching
$I_{vv'}^0$ is reduced by the Stern-Volmer factor \Cite{Stern-Volmer}
\begin{equation}
\Label{eq:I_q} I_{vv'}(P) = I_{vv'}^0\frac{1}{1+P/P'_v}\,.
\end{equation}

From \eref{eq:Rel_I_0} and \eref{eq:I_q} the relative intensities of
molecular bands at high pressure ($P\gg P'$)
%%FF
become~\footnote{Experimental confirmation of a collisional mechanism populating vibrational levels of the C$^3\Pi_u$ state
which might induce an additional pressure dependence of the 2P relative intensities has been shown in
\Cite{5thFW_Morozov}\Cite{Morozov_08}.}
%%FF

\begin{equation}
\Label{eq:Rel_I}
 \frac{I_{vv'}}{I_{00}}=\frac{q_{X\rightarrow v}}{q_{X\rightarrow 0}}
  \frac{B^{vv'}}{B^{00}}\frac{P'_v}{P'_0}\,.
\end{equation}
Collisional quenching enlarges the total transition probability and,
therefore, the lifetime of the population of excited molecules
$\tau_v$ is shortened as compared with the radiative one as
\begin{equation}
\Label{eq:tau_v} \frac{1}{\tau_v}=\frac{1}{\tau^r_v}+\frac{1}{\tau^c_v}\,,
\end{equation}
with $\tau^c_v=1/A^c_v$. As a result, the effective lifetime decreases
with pressure as
\begin{equation}
 \Label{eq:tau_SV}
 \frac{1}{\tau_v(P)}=\frac{1}{\tau_v^r}\left(1+\frac{P}{P'_v}\right)\,.
\end{equation}

Both $\tau^r_v$ and $P'_v$ can be measured in a plot of reciprocal
lifetime versus pressure (Stern-Volmer plot). This is a very well
established technique in use since many years for the experimental
determination of radiative lifetimes and quenching cross sections. In
principle, a measure of the fluorescence intensity versus pressure
\eref{eq:I_q} also provides a determination of $P'_v$. However, as
discussed in detail later, in a laboratory experiment the above
relationship might be distorted because of the effect of secondary
electrons leading to systematic uncertainties in the measurement of
the characteristic pressures.

The probability of collisional de-excitation per unit time of a
molecule in a given upper level $v$ can be expressed\footnote{In the
next paragraphs all collisional parameters will be assumed to
correspond to a given molecular level $v$.} in terms of the quenching
rate constant $K_Q$ as $A_c=NK_Q$ where $N$ is the number of molecules
per unit volume. In the case of pure nitrogen
\begin{equation}
\Label{eq:K_Q} K_Q =  \sigma_{{\rm NN}} \bar{v}
\end{equation}
and
\begin{equation}
\Label{eq:v_bar} \bar{v}=\sqrt{\frac{16kT}{\pi M}}\,,
\end{equation}
where $\sigma_{{\rm NN}}$ is the cross section for collisional de-excitation
between nitrogen molecules, $\bar{v}$ is the mean value of the relative
velocity of molecules in the gas, $T$ is the absolute temperature, $M$
is the molecular nitrogen mass, and $k$ is the Boltzmann's constant.
Since $P=NkT$, the characteristic pressure for pure nitrogen can be
expressed as
\begin{equation}
 \Label{eq:P_N} P'_{\rm N} = \frac{kT}{\tau}\frac{1}{\sigma_{{\rm NN}} \bar{v}} =
  \frac{\sqrt{\pi MkT}}{4\sigma_{{\rm NN}}} \frac{1}{\tau^r}\,.
\end{equation}

The above expressions can be generalized for a mixture of gases as
\begin{equation}
\Label{eq:1_P} \frac{1}{P'} = \sum_i{\frac{f_i}{P'_i}}\,,
\end{equation}
where $f_i$ is the fraction of molecules of type $i$ in the mixture and
\begin{equation}
\Label{eq:P_i} P'_i = \frac{kT}{\tau}\frac{1}{\sigma_{{\rm N}i} \bar{v}_{{\rm N}i}}\,.
\end{equation}
In the general case, the relative velocity $\bar{v}_{{\rm N}i}$ is given 
by \Cite{Yardley}
\begin{equation}
\Label{eq:rel_v_bar} \bar{v}_{{\rm N}i}=\sqrt{\frac{8kT}{\pi \mu}}\,,
\end{equation}
where $\mu = M_{\rm N} M_i /(M_{\rm N}+M_i)$ is the reduced mass of
the two body system N-$i$.

For dry air the above sum includes basically nitrogen and oxygen with
$f_{\rm N}$ = 0.79 and $f_{\rm O}$ = 0.21. However, in practice, air
contains also other components. For instance, the effect of argon can
be treated by \eref{eq:1_P} and \eref{eq:P_i} accordingly.

A particular interesting case is the effect of water vapor. The
characteristic pressure of humid air $P'_{hum}$ containing a fraction
$f_{{\rm H}_2{\rm O}}$ of water molecules, that is a water vapor
pressure $P_{{\rm H}_2{\rm O}}$ = $Pf_{{\rm H}_2{\rm O}}$, is related
with that of dry air $P'_{dry}$ by
\begin{equation}
\Label{eq:P_hum} \frac{1}{P'_{hum}} = \frac{1}{P'_{dry}}\left(1 -
\frac{P_{{\rm H}_2{\rm O}}}{P}\right)+
\frac{P_{{\rm H}_2{\rm O}}}{P}\frac{1}{P'_{{\rm H}_2{\rm O}}}\,.
\end{equation}

Laboratory measurements of $P'$ for nitrogen with variable quantities
of water vapor, argon, oxygen, etc.~provide values of the
corresponding $P'_i$ pressures and, therefore, the dependence of
fluorescence intensity on environmental conditions.

Quenching collision is a very complex problem of molecular physics and
basically no reliable theoretical predictions on cross sections are
available.  Therefore, a description of the dependence of fluorescence
quenching on  pressure and humidity relies on experimentally
determinated $P'$ values.

Furthermore, fluorescence quenching depends on temperature for a given
density and air composition. Firstly, the frequency of collisions
increases with $\bar{v}$ and, thus, $P'$ grows with $\sqrt{T}$
\eref{eq:P_N}. Secondly, the collisional cross section is a function
of the kinetic energy of the colliding particles  and, thus,
$\sigma_{{\rm N}i}$ is a function of $T$. While the first dependence
obeys well known laws of the kinetic theory of gases, the second one
is again associated to the molecular problem of collisional
de-excitation, in this case on the dependence of the quenching cross
section on collision energy.  Very few experimental studies of the
temperature dependence for nitrogen fluorescence are available. On the
other hand, no simple theory has been developed capable to predict the
temperature dependence of quenching. The collisional cross section is
assumed to follow a power law in temperature\footnote{See
\Cite{5thFW_Fraga}, \Cite{T_power_law}, and \Cite{Bailey} for
discussions on the $T$ dependence of the quenching cross section.},
$\sigma \propto T^{\alpha}$, where the $\alpha$-parameter might be
either positive or negative, depending on the nature of the partners
and the type of interaction. The $\alpha$-coefficient might even be
valid only for certain temperature ranges, since the dominating type
of interaction varies with the velocity of the molecules. As a
consequence, from \eref{eq:I_q} a dependence of fluorescence intensity
as
\begin{equation}
\Label{eq:1_Ia} \frac{1}{I}  \propto 1+ b\:T^{\alpha-\frac{1}{2}}\,,
\end{equation}
can be predicted for a temperature scan at constant pressure, while
the dependence at constant gas density $\rho=P/(R_{gas}T)$ ($R_{gas}$
is the specific gas constant) follows
\begin{equation}
\Label{eq:1_Ib} \frac{1}{I}  \propto 1+ b'T^{\alpha+\frac{1}{2}}\,,
\end{equation}
where $b$ and $b'$ are constants.

In this volume new interesting measurements of the $T$ dependence for
pure nitrogen \Cite{5thFW_Fraga} and air \Cite{5thFW_airfly_T} will be
presented.

\subsection{Fluorescence yield}
\Label{sec:yield}
Several parameters can be used to quantify the intensity of
air-fluorescence radiation in regard with the energy deposited by
electrons. In addition, the same physical magnitudes are denominated
in  the literature with different names and/or using different
symbols.

The main physical magnitudes are the following:
\begin{itemize}
\item Number of fluorescence photons emitted per unit electron path
length.  Several authors (for instance \Cite{arqueros_Astr_Ph}%
\Cite{kakimoto1996}\Cite{nagano2003}\Cite{nagano2004}\Cite{flash_06})
name it fluorescence (or photon) yield.
\item Fraction of deposited energy emitted as fluorescence radiation
(without units) named fluorescence efficiency in
\Cite{kakimoto1996}\Cite{nagano2003}\Cite{nagano2004}.
\item Number of fluorescence photons emitted per unit deposited
energy. For several authors (for instance \Cite{airfly_P}%
\Cite{flash_07}\Cite{5thFW_Waldenmaier}) this parameter is the
fluorescence yield.
\end{itemize}

In this article we will use the following definitions and symbols:
\begin{itemize}
\item $\varepsilon_{\lambda}$ [m$^{-1}$] is the number of fluorescence
photons with wavelength $\lambda$ corresponding to a given transition
$v-v'$
\footnote{In the next paragraphs, until the end of this
section, the wavelength $\lambda$ will characterize the molecular
transition instead of the $vv'$ pair. This is a more compact notation
very common in articles of the air-fluorescence community.} 
per unit of electron path length.
\item $\Phi_{\lambda}$ is the fluorescence efficiency, defined as the
fraction of deposited energy emitted as fluorescence radiation.
\item The fluorescence yield $Y_{\lambda}$ [MeV$^{-1}$] is defined as
the  number of fluorescence photons emitted per unit deposited
energy.
\end{itemize}

The ratio of $\Phi_{\lambda}$ and $Y_{\lambda}$ is easily given by the
energy of the photons $E_{\lambda} = h\nu$ with $\nu =c/\lambda$.
However, the relationship between $\varepsilon_{\lambda}$ and
$Y_{\lambda}$ is not straightforward. As discussed below, fluorescence
light is basically generated by secondary electrons produced in
ionization processes. These secondaries have a non-negligible range
and, therefore, measured fluorescence intensity depends on geometrical
features of the observation volume. A precise measurement of the
fluorescence yield requires the evaluation of deposited energy in the
same gas volume from where fluorescence is being detected.

The total number $\varepsilon_\lambda$ of fluorescence photons generated
per unit path length in a very large medium can be expressed as a
function of the optical cross section $\sigma_{\lambda}$ of the
transition by \Cite{arqueros_Astr_Ph}
\begin{equation}
\Label{eq:varepsilon_total} \varepsilon_{\lambda} =
N\frac{\sigma_{\lambda}}{1+P/P'_{\lambda}}\,,
\end{equation}
where $N$ is the density of nitrogen molecules and
$P'_{\lambda}$ is the characteristic pressure of the upper level $v$
of the transition.  Obviously, $\varepsilon_{\lambda}$ depends on
electron energy because of the energy dependence of the optical cross
section.

In a laboratory experiment with a finite observation volume, a
fraction\footnote{This fraction depends on the gas pressure and the
geometrical features of the experimental set-up.} of these photons is
not detected by the system.  \Eref{eq:varepsilon_total} can be applied
using an effective optical cross section $\sigma_{\lambda}^{eff}$ as
defined in \Cite{blanco}.

The number of fluorescence photons per unit column density
per electron is given by
\begin{equation}
\Label{eq:varepsilon_rho} \frac{\varepsilon_{\lambda}}{\rho} =
\frac{N_A}{M_{gas}}\frac{\sigma^{eff}_{\lambda}}{1+P/P'_{\lambda}}=
\frac{A_{\lambda}}{1+P/P'_{\lambda}}\,,
\end{equation}
where $\rho$ is the gas density, $N_A$ is Avogadro's number and
$M_{gas}$ is the mass of the gas molecules. The value of
$\varepsilon_{\lambda}/\rho$ in the absence of quenching ($P=0$) is
named $A_{\lambda}$ in \Cite{kakimoto1996}\Cite{nagano2003}.

The fluorescence efficiency depends on pressure as
\begin{equation}
 \Label{eq:efficiency}
 \Phi_{\lambda} = \Phi_{\lambda}^0 \frac{1}{1+P/P'_{\lambda}}\,.
\end{equation}
At zero pressure, $\Phi_{\lambda}$ is given by
\begin{equation}
\Label{eq:efficiency_0}
  \Phi_{\lambda}^0 =
       \frac{\rho \, A_{\lambda}\,h\nu}{({\rm d}E/{\rm d}X)_{dep}}\,,
\end{equation}
where $({\rm d}E/{\rm d}X)_{dep}$ is the energy deposited per unit
electron path length in the same volume where fluorescence photons
have been generated (see below for more details on the effect of
secondary electrons).

Finally, the fluorescence yield follows the same pressure dependence
as $\Phi_\lambda$
\begin{equation}
\Label{eq:FY} Y_{\lambda} = Y_{\lambda}^0 \frac{1}{1+P/P'_{\lambda}}\,,
\end{equation}
and its value in the absence of quenching is given by
\begin{equation}
\Label{eq:FY_0} Y_{\lambda}^0 = \frac{\Phi_{\lambda}^0}{h\nu}\,.
\end{equation}
\Erefs{eq:FY} and \eref{eq:FY_0} can be written as
\Cite{kakimoto1996}\Cite{nagano2003}
\begin{equation}
\Label{eq:FY_N} Y_{\lambda} = \frac{1}{({\rm d}E/{\rm d}X)_{dep}}\frac{\rho
A_{\lambda}}{1+\rho B_{\lambda}\sqrt{T}}\,,
\end{equation}
where
\begin{equation}
\Label{eq:B_l_N} B_{\lambda} = \frac{R_{gas}\sqrt{T}}{P'_{\lambda}} .
\end{equation}

The dependence of the fluorescence yield on pressure, temperature,
humidity, etc. can be predicted from either \eref{eq:FY} and
\eref{eq:FY_0} or \eref{eq:FY_N}, using the characteristic pressure as
given above in \eref{eq:1_P} and \eref{eq:P_i}.

\section{The actual status}
\Label{sec:status}

This section describes the progress in the last few years on the
experimental and theoretical tools developed for air-fluorescence
studies. In \sref{sec:experimental_techniques} the modern experimental
techniques used for the measurement of fluorescence yield and its
dependence on atmospheric parameters are described. Electron sources,
target features, detection systems as well as the various techniques
developed for the absolute calibration of the systems are described.
Finally, in \sref{sec:fl_methods} theoretical results on the processes
leading to the emission of air fluorescence light and the relation to
deposited energy are discussed.

\subsection{Experimental techniques}
\Label{sec:experimental_techniques}
Any experimental set-up consists basically of three components: a
source of electrons (or $\alpha$-particles) properly monitored, a
collision chamber where air or any gas mixture is excited by the
electrons, and an optical as well as an electronic system to register
the fluorescence light intensity.

\subsubsection{Electron sources}
\Label{sec:electron_sources}
Three types of sources are used in air-fluorescence experiments:
electron beams from accelerators in large facilities, radioactive
sources, and low-energy electron guns in laboratories.

\paragraph{Accelerators}
They can provide electron beams with a small diameter typically of
about few millimeters. Different kinds of accelerators are available
for the various energy ranges (keV -- GeV). In particular, they are
the only possible source for very high-energy electrons. A
disadvantage of this technique is the large background signal induced
in the fluorescence detectors which requires a careful subtraction
from the fluorescence signal. Furthermore, electrons exit the
accelerator line through a window of a certain thickness and material
dependent on the energy range.

The FLASH Collaboration
\Cite{flash_06}\Cite{flash_07}\Cite{5thFW_FLASH_thin} used the Final
Focus Test Beam facility at the Stanford Linear Accelerator Center
which provided 28.5~GeV electrons in 3~ps pulses of about 10$^8$
electrons at a rate of 10~Hz.

The MACFLY Collaboration \Cite{MACFLY_thin} used the CERN/SPS-X5
test beam facility which delivers a pulsed electron beam of about
10$^4$ electrons per spill (4.8~s duration) every 16.8~s. Measurements
at 20 and 50 GeV were carried out using this facility.

The AIRFLY Collaboration exploits this technique in an ambitious
program to measure the fluorescence yield in the interval 6~keV --
420~MeV using four different accelerators~\Cite{5thFW_airfly_E}. The
interval 6 -- 30~keV is covered by the Advanced Photon Source of the
Argonne National Laboratory.  The intense synchrotron x-ray beam of
the 15-ID line of this accelerator produces an almost monochromatic
beam of electrons through photoelectric and Compton interactions with
the ambient air. Also at the Argonne National Laboratory, the
Chemistry Division electron Van de Graaff accelerator operated
in pulsed mode at 60~Hz, with beam currents from 0.2 to 0.8~$\mu$A,
was used by this collaboration to get electrons in the range 0.5
-- 3 MeV.  Measurements in the energy range from 3 to 15~MeV were
performed at the Argonne Wakefield Accelerator. The LINAC was operated
at 5~Hz, with bunches of maximum charge of 1~nC, length 15~ps
(FWHM), and a typical  energy spread of $\pm$0.3~MeV at 14~MeV. Finally,
measurements in the energy region 50 -- 420~MeV were performed by
AIRFLY at the Beam Test Facility of the INFN Laboratori Nazionali di
Frascati, which can deliver electrons with intensity ranging from
single particle up to about 100 particles per bunch at a repetition
rate of 50~Hz with a typical pulse duration of 10~ns.

\paragraph{Radioactive sources}
Beta emitters provide electrons with a continuous energy spectrum. In
particular, $^{90}$Sr-$^{90}$Y sources with a maximum and average
energy of 2.3 and 0.85~MeV, respectively, are widely used. This energy
range, around the minimum of the energy loss curve, is of great
interest in air-fluorescence studies. In this technique fluorescence
detection is based on electron-photon coincidences.

\begin{figure}[t]
\centering
\epsfig{width=0.49\textwidth, file=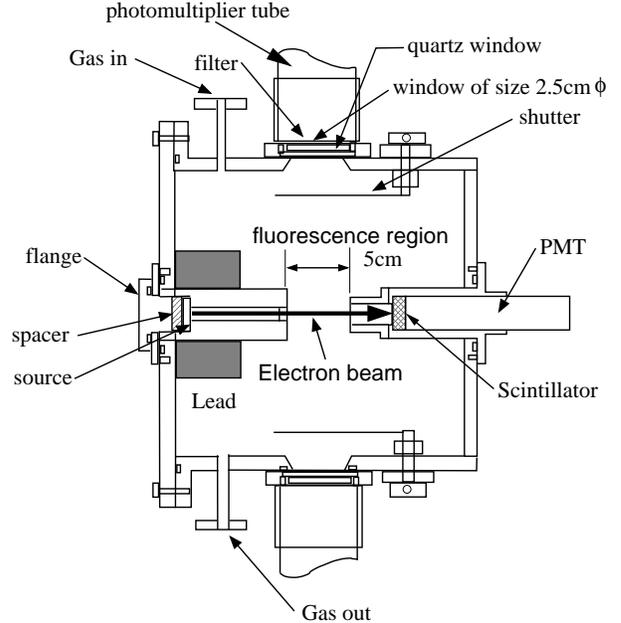}
\caption{Schematic drawing of the chamber (top view) used by Nagano et
al.  \Cite{nagano2003}. Three photomultipliers are mounted on two
sides and the top of the chamber, they view fluorescence light through
quartz windows. Optical filters are mounted between the
photomultipliers and the windows.  Electrons from a $^{90}$Sr
radioactive source are beamed and detected by a scintillation
counter.}
\Label{fig1_sakaki}
\end{figure}

A schematic drawing of the chamber used by Nagano et al., applying
this technique, is presented in \fref{fig1_sakaki} \Cite{nagano2003}.

An advantage of using radioactive sources is that once the source is
safely located in the experimental set-up, long lasting experiments
can be carried out at very low maintenance costs. However, this
technique has also some disadvantages. The main one is that unless
strong radioactive sources are used, the rate of coincidences is very
low and, thus, very large data acquisition times are needed to achieve
sufficient statistics, then increasing systematic uncertainties.

In the last years, sources with increasing activity have been used.
Nagano et al. \Cite{nagano2003}\Cite{nagano2004}, using a 3.7 MBq
$^{90}$Sr source, were able to measure the absolute value and the
pressure dependence of fluorescence for isolated bands using broadband
10 nm filters. A source of 37 MBq has allowed Waldenmaier et al.
\Cite{5thFW_Waldenmaier} (AIRLIGHT experiment) accurate measurements
of the very short nitrogen lifetimes in air at high pressure. The most
active source ever used in this field is the one of Lefeuvre et al.
\Cite{Paris} with an activity of 370 MBq which has allowed to record
the spectrum with a monochromator of 6 nm resolution.  Finally, the
MACFLY Collaboration also used a $^{90}$Sr source for measurements of
the absolute fluorescence yield at 1.5 MeV \Cite{MACFLY_thin}.

Alpha emitters are also a very useful tool for fluorescence studies.
As an example, an interesting study on pure nitrogen using a
$^{241}$Am source of 3.7 kBq has been carried out \Cite{5thFW_Fraga}.
Alpha particles lose energy by excitation and ionization in the gas.
Although direct excitation of the N$_2$ 2P system is forbidden,
low-energy secondary electrons excite the C$^3\Pi_u$ upper level (see
\Cite{5thFW_Fraga} for more details).
Many important properties of air fluorescence like its dependence on
pressure, temperature, and humidity can be studied using
this technique.  Again, the main disadvantage is the low
fluorescence intensity due to the limited source activity which in
this case might be more important for legal restrictions of alpha
sources.

\paragraph{Low-energy electron guns} Air-fluorescence emission
induced by low-energy electrons ($E< 0.1$ MeV) is of great interest.
In the first place, a non-negligible fraction of the energy deposited
in the atmosphere by a cosmic-ray shower is delivered by low-energy
electrons \Cite{risseheck}. Furthermore, the assumption of
proportionality between fluorescence intensity and deposited energy
might not be fulfilled at low electron energies \Cite{5thFW_Arqueros}.

Customized electron guns are being used for this application.
Morozov et al.~\Cite{morozov2005} employ a modified electron gun
designed for monochrome displays. The cathode is operated at high
negative potential and the anode is connected to the ground. The gun
delivers electrons of about 12 keV. Electron pulses of about 5 ns FWHM
are performed by means of the control grid of the electron gun.

Rosado et al. \Cite{5thFW_Rosado} have designed a novel gun.  The
electron emission is based on a plasma produced by a pulsed nitrogen
laser focused on the cathode, with up to 30 Hz repetition rate. The
cathode, which is maintained at a negative potential by a high voltage
power supply, accelerates electrons to kinetic energies up to 30 keV.
Electron pulses of about 20 ns width and 40 mA peak intensity are
achieved using this technique.

For fluorescence studies at high pressure induced by low-energy
electrons, a very thin window has to be used to isolate the electron
gun from the collision chamber. Morozov et al. use an ultra thin (300
nm) silicon nitride window which allows the passage of 12~keV
electrons without substantial energy degradation.  For the moment,
Rosado et al.~are working at low pressure.  Differential pumping
allows maintaining pressure in the electron gun well below 0.1 Pa to
ensure cathode isolation, whereas working pressures up to 35 Pa can be
used in the gas cell.

\subsubsection{The target}
\Label{sec:target}

Concerning  the target two types of fluorescence experiments are being
carried out. The so-called thick-target and thin-target experiments.
For high-energy electrons ($E \gtrsim $ 1~MeV), air can be considered
a thin target since the attenuation of the beam is very small even at
atmospheric pressure.  Most experiments described here are carried out
under thin target conditions  (e.g.\ the Nagano experiment shown in
\fref{fig1_sakaki}).

In our field thick-target experiments are those which use a dense
medium where the high-energy electrons initiate an electromagnetic
shower which enters the air collision chamber to produce fluorescence
light. In other words, in these thick-target experiments the
fluorescence light produced by an electromagnetic shower (created in a
thick non-air target) on a thin air target is studied.

\begin{figure}[t]
 \centering
 \epsfig{width=0.49\textwidth, file=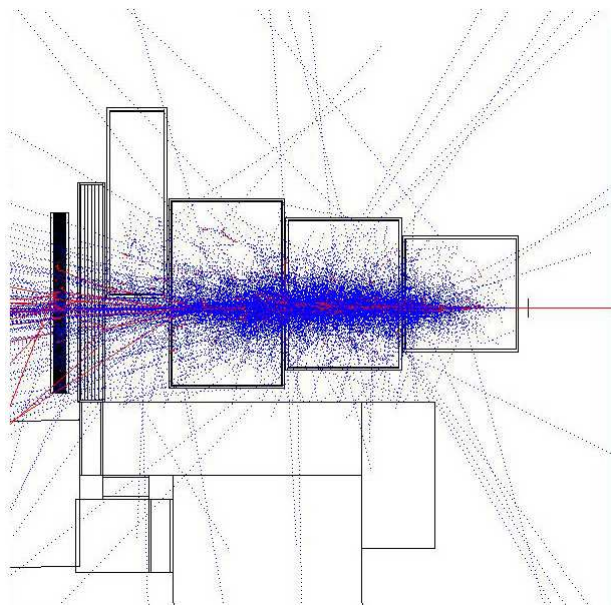}
 \caption{Thick-target configuration of the FLASH Collaboration showing a
      GEANT3.2 simulation of an electromagnetic shower generated in
      the target \Cite{5thFW_FLASH_thick}\Cite{flash_thick}.}
 \Label{fig5_flash_thick}
\end{figure}

Two experiments have used the thick-target technique, FLASH and
MACFLY.  A schematic view of the thick-target  configuration of the
FLASH Collaboration is displayed in \fref{fig5_flash_thick}.  The result
of a GEANT3.2 simulation of an electromagnetic shower generated in the
target is shown \Cite{5thFW_FLASH_thick}\Cite{flash_thick}.  The
electron beam is incident on a variable-thickness ceramic alumina
stack.  This material has good thermal properties along with
possessing a critical energy similar to that of air.

The MACFLY thick device \Cite{MACFLY_thick} is composed of an
internally black covered quasi cylindrical, large volume ($\approx$
1~m$^3$), pressurized tank containing the gas under study. The electron
beam, aligned with the axial symmetry of the chamber, is impinging on
a pre-shower target.  This variable thickness pre-shower system, is
used to initiate electromagnetic showers inside the chamber.

A particular case is an air target at very low pressure ($<$ 35 hPa)
for the study of the fluorescence contribution of secondary electrons
as used by Rosado et al.~\Cite{5thFW_Rosado}.  The experimental
conditions, i.e.\ low pressure and low energy ($\sim$ 30 keV) are
suitable for the analysis of spatial features of the fluorescence
emission. This experiment has allowed to test the results of a
model~\Cite{5thFW_Arqueros}, discussed below in \sref{sec:Pred_Fluor},
for the calculation of the fluorescence light generated by secondary
electrons.

The accurate knowledge of the dependence of fluorescence intensity on
environmental conditions is one of the most important goals in this
field. The gas target where fluorescence is produced in the laboratory
is in general a mixture of gases emulating air under various
atmospheric conditions. Since air-fluorescence light is basically
produced by nitrogen, many experiments have been performed using pure
nitrogen as target.  Fluorescence in air is strongly quenched by
oxygen collisions while pure nitrogen is much more efficient. Thus,
properties of N$_2$ fluorescence can be more easily studied using pure
nitrogen.

\begin{figure}[t]
 \centering
 \epsfig{width=0.49\textwidth, file=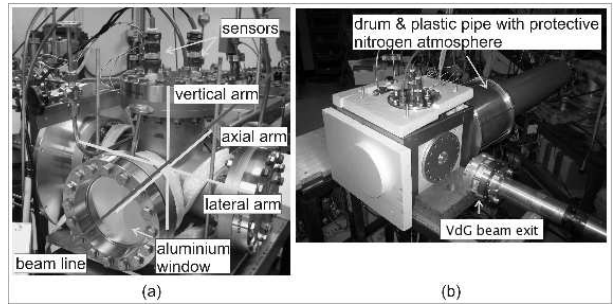}
  \caption{The temperature chamber used by the AIRFLY Collaboration
       \Cite{5thFW_airfly_T}: a) the chamber at the beam line as
       it appears before mounting the polystyrene box, b) the
       chamber inside the polystyrene box, with a protective drum
           and pipe also in place.}
 \Label{fig1_AIRFLY_T}
\end{figure}

Several experiments have been carried out to check the effect of argon
on air fluorescence (e.g.~\Cite{5thFW_airfly_P}).  The effect of
humidity has been studied as well
\Cite{5thFW_airfly_T}\Cite{flash_07}\Cite{5thFW_Waldenmaier}\Cite{morozov2005}\Cite{5thFW_nagano},
adding to the mixture a known amount of water vapor.  Another very
important parameter is the air temperature. The dependence of
fluorescence yield on temperature in a large interval, covering that
found in the atmosphere, is an experimental challenge.  Devices
capable to provide air targets under controlled temperatures in the
required interval have been designed by AIRFLY \Cite{5thFW_airfly_T}
(\fref{fig1_AIRFLY_T}) and Fraga et al.  \Cite{5thFW_Fraga}. See these
articles for details on the chamber design.

\subsubsection{Detection techniques}
\Label{sec:det_techn}

The detection and analysis of the air-fluorescence radiation are
carried out using the appropriate optical and electronic devices. In
the first place, the emitted fluorescence light has to be collected
and, if possible, spectroscopically analyzed.  For the latter task, a
set of filters or a monochromator are used. For interference filters
the dependence of the spectral response on the incident angle has to
be carefully measured
\Cite{5thFW_Fraga}\Cite{5thFW_Waldenmaier}\Cite{5thFW_airfly_A}.  If
sufficient fluorescence intensity is available, a monochromator can be
used to measure the fluorescence spectrum
\Cite{5thFW_airfly_P}\Cite{Paris}\Cite{5thFW_Rosado}.  The spectrum
provides a measure of the relative intensities which is a very
valuable information. In addition, the dependence on pressure of
either reciprocal lifetimes \Cite{5thFW_Waldenmaier} or intensities
\Cite{5thFW_airfly_P}\Cite{nagano2004} of spectroscopically resolved
fluorescence yield allows to measure the $P'_v$ values.

For the detection of fluorescence light, the most usual tool is a
photomultiplier working in single photon counting regime. In fact, if
possible, several photomultipliers viewing the collision chamber from
several viewpoints allow a higher efficiency and the possibility to
record simultaneously the fluorescence radiation in different spectral
intervals. The AIRFLY Collaboration uses additionally a hybrid
photodiode capable of single photoelectron counting
\Cite{5thFW_airfly_A}.

\subsubsection{Absolute calibration}
\Label{sec:abs_calibration}

The most important objective of this world-wide effort is an accurate
measurement of the absolute value of the air-fluorescence yield. For
this task, it is necessary to calibrate the detection system
absolutely, including geometrical and transmission factors of the
entire optical system.  The absolute value of the number of electrons
traversing the field of view of the collision chamber has to be
measured.

To determine the fluorescence yield, the energy deposited by the
electron beam inside the volume observed by the optical system has to
be known. A first approach, which might be valid at low electron
energy is to assume that the energy loss, as predicted by the
Bethe-Bloch formula, equals the energy deposited by the electrons in
the medium \Cite{kakimoto1996}\Cite{nagano2003}.

As discussed later, secondary electrons generated by the primary
electron are mainly responsible of both the fluorescence emission and
the energy deposition in the medium. Therefore, the total size of the
volume where energy is deposited (and fluorescence is emitted) is
related to the range of secondary electrons.  For high-energy
primaries, a non-negligible fraction of the energy is deposited by
secondaries with a range larger than the typical size of the
experimental collision volume observed by the optical system.  In this
case, a Monte Carlo simulation is very useful to determine the energy
deposited in the interaction region accurately, including the
geometrical features of the collision chamber and the optical field of
view.  Several standard Monte Carlo codes, like EGS4 \Cite{EGS4} and
GEANT4 \Cite{GEANT4} are being used for this purpose.

Several techniques have been developed for the absolute calibration of
the optical systems. In principle, an accurate measurement of the
geometrical features of the electron beam, the collection system, the
transmission of all the optical elements, and the quantum efficiency
of the light detector provide the necessary efficiency factor. This
procedure has been applied by several experiments
\Cite{nagano2004}\Cite{5thFW_Waldenmaier}\Cite{MACFLY_thin}\Cite{Paris}.

Other calibration procedures have been developed in order to reduce
the (usually large) systematic uncertainties from the efficiency
parameters mentioned above.  These techniques rely on the comparison
of fluorescence intensity with a well-known physical process leading
to the emission of light with the same spectral and geometrical
features.  Two physical processes have been employed for this purpose.
The first one uses the \Cerenkov light emitted by the electron beam in
the gas, while the second one is based on Rayleigh scattering from a
laser beam replacing the electron beam.

\begin{figure}[t] \centering
 \epsfig{width=0.49\textwidth, file=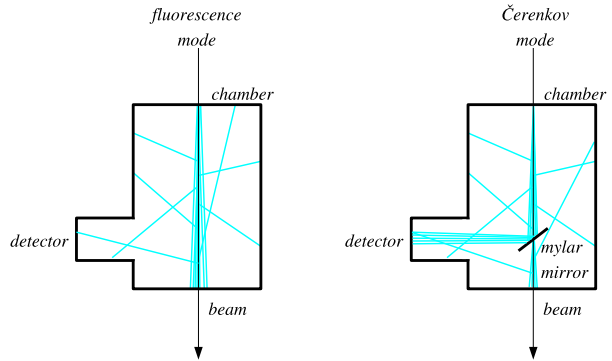}
  \caption{Experimental set-up used by the AIRFLY Collaboration for
           the measurement of the absolute air-fluorescence yield
           \Cite{5thFW_airfly_A}.}
 \Label{fig1_AIRFLY_Absolute}
\end{figure}

The AIRFLY Collaboration \Cite{5thFW_airfly_A} has developed the
technique based on the comparison with \Cerenkov light. The
measurements are taken in two modes (\fref{fig1_AIRFLY_Absolute}). In
the fluorescence mode, the isotropic fluorescence light produced by
the electrons in the field of view of the detector is recorded. In
this mode, contributions from other sources of light, like \Cerenkov
or transition radiation, are negligible due to the non-isotropic
emission of such mechanisms. In the \Cerenkov mode, a thin mylar mirror
at an angle of 45$^\circ$ is inserted remotely into the beam,
redirecting the \Cerenkov light into the detector. In this mode, the
\Cerenkov light fully dominates over fluorescence.
The absolute fluorescence yield is then determined using the ratio of
the signal measured in the fluorescence and in the \Cerenkov
configurations. The \Cerenkov yield is known from theory, the
geometrical factors of the apparatus are derived from a full GEANT4
simulation of the detector and take into account the probability of a
photon being emitted in each case and also the fact that \Cerenkov
light is very directional and fluorescence light is emitted
isotropically.  Using this technique, the AIRFLY Collaboration has
measured the absolute yield of the 337 nm band. A preliminary result
has been presented already in \Cite{5thFW_airfly_A}.

The technique based on a comparison with the Rayleigh-scattered light
was proposed by the FLASH and the Madrid groups at a previous workshop
\Cite{IWFM05}. A nitrogen pulsed laser beam crosses the collision
chamber in the place of the electron beam.  Typical pulses of about
100 $\mu$J energy and 4 ns width scatter a number of photons of the
same order of magnitude as those from fluorescence runs. One of the
main problems of this technique is light scattered at the walls of the
chamber which has to be carefully suppressed. A measurement of the
pressure dependence of the Rayleigh signal provides valuable
information on the background scattered light and the linearity of the
signal.

Using this technique, the FLASH Collaboration has already carried out
a measurement of the absolute yield with an uncertainty below 10\%.
The Madrid \Cite{5thFW_Rosado} and AIRLIGHT \Cite{5thFW_Waldenmaier}
groups are presently using this technique for the absolute calibration
of their systems.

Two different strategies are being used for the measurement of the
total fluorescence yield in the spectral interval of the telescopes.
Several experiments \Cite{nagano2004}\Cite{MACFLY_thin}\Cite{Paris}
carry out an absolute measurement of the yield for the whole spectral
interval including many molecular bands while \Cite{5thFW_airfly_A}
measures the absolute fluorescence yield of the main band (337nm) and
the contribution of the remaining spectral components is inferred
later from an accurate measurement of the spectrum
\Cite{5thFW_airfly_P}.

Notice that a comparison of results of the fluorescence yield in
different spectral intervals needs a value for the relative
intensities as well as the $P'_v$ values, see \eref{eq:Rel_I}.

\subsection{Theoretical approaches}
\Label{sec:fl_methods}

\subsubsection{Predictions on fluorescence emission}
\Label{sec:Pred_Fluor}

The well-known physical processes leading to molecular excitation and
fluorescence emission have been described in \sref{sec:fl_physics}.
Einstein coefficients, Franck-Condon factors as well as excitation and
ionization cross sections are available in the literature, e.g.\
\Cite{gilmore}\Cite{laux}. The amount of fluorescence photons
generated by electrons traversing a given air thickness has been
calculated in \Cite{blanco}\Cite{arqueros_Astr_Ph} using a Monte Carlo
algorithm which takes into account the dominant role of secondary
electrons.  This algorithm also calculates the energy deposition for
which secondary electrons are mainly responsible.

Bunner \Cite{bunner1967} realized early that secondary electrons from
ionization processes are the main source of fluorescence light, since
the excitation cross section of the corresponding upper levels
(\fref{exc_cros_sect}) shows a fast decrease with energy, in
particular the one for the 2P system. Unfortunately, Bunner
\Cite{bunner1967} was not able to calculate the fluorescence emission
from secondary electrons since the necessary data, in particular, the
spectrum of secondary electrons were not available at that time for
collisions at high energy. An estimate of the energy spectrum of
secondary electrons up to the GeV range has been used in
\Cite{blanco}\Cite{arqueros_Astr_Ph} to calculate for the first time
the fluorescence intensity induced by high-energy electrons. An
improved energy spectrum of secondaries has allowed recently more
reliable results of the fluorescence intensity and also a precise
calculation of deposited energy and, thus, of the fluorescence yield
\Cite{5thFW_Arqueros}.

\begin{figure}[t] \centering
 \epsfig{width=0.49\textwidth, file=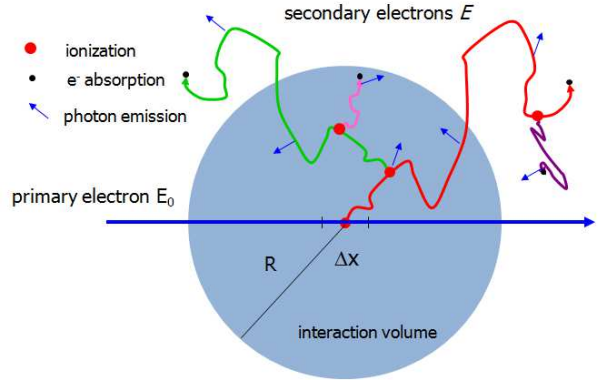}
 \caption{The energy loss of a primary electron in $\Delta X$ gives
      rise to the production of secondaries, being mainly
      responsible for the fluorescence light emission. A fraction of
      both deposited energy and fluorescence emission might take
          place outside the observation region
          \Cite{5thFW_Arqueros}.}
 \Label{processes}
\end{figure}

A schematic view of the processes involved in the emission of
fluorescence light and the deposition of energy from secondary
electrons as modeled in \Cite{5thFW_Arqueros} can be seen in
\fref{processes}. A primary electron traversing an atmospheric depth
$\Delta X$ may either excite or ionize a molecule. In the latter case,
the secondary electron produces further excitations and/or ionizations
until all secondaries are stopped in the medium. Both fluorescence
generation and energy deposition due to molecular
excitations/ionizations are calculated using a Monte Carlo algorithm.
As a result, the energy deposited per unit path length of air as well
as the number of molecules excited to the upper levels of the 2P and
1N system are determined. The results for both magnitudes depend on
the volume of the interaction region as well as the air pressure. In
fact, it is a function of $P\times R$, where $R$ is the radius of the
sphere around the interaction point defining the medium size.
Neglecting the quenching effect, the ratio of both magnitudes gives
$Y^0$, i.e.\ the fluorescence yield at $P$=0.  Results on these
predictions are compared with experimental data in
\Cite{5thFW_Arqueros}.

\begin{figure}[t] \centering
 \epsfig{width=0.49\textwidth, file=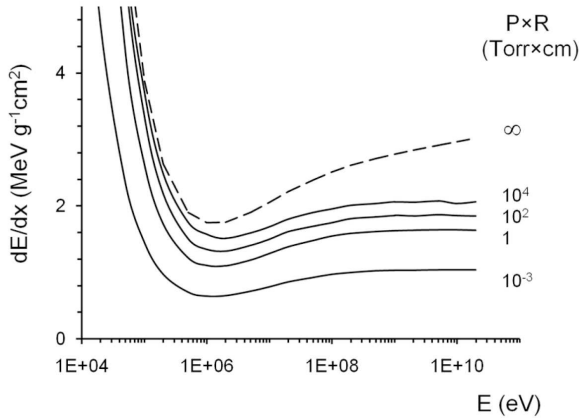}
 \caption{Energy deposited by a primary electron per unit path length versus
 primary energy for several values of $P\times R$ (product of pressure times
 fiducial region, continuous lines). For very high $P\times R$ values deposited
 energy equals the energy loss of the primary electron predicted by the
 Bethe-Bloch theory (broken line) \Cite{5thFW_Arqueros}.}
 \Label{fig5_arqueros}
\end{figure}

The predicted values of the energy deposited per unit path length as a
function of electron energy for several values of $P\times R$ are
depicted in \fref{fig5_arqueros} \Cite{5thFW_Arqueros}.  As expected,
deposited energy at very high $P\times R$ values tends to the total
energy loss predicted by the Bethe-Bloch theory.  Notice that for
typical observation volumes in fluorescence experiments ($\approx
10^3$ hPa$\times$cm), the deposited energy is smaller than the total
energy loss by an amount which at atmospheric pressure ranges from
about 18\% at 1 MeV to 42\% at 10 GeV.

\subsubsection{The relation between fluorescence light intensity
and deposited energy}
\Label{sec:Edep}

The fluorescence technique is based on the assumption that the
fluorescence yield is independent of the electron energy. In other
words, for given atmospheric conditions the fluorescence intensity is
assumed to be proportional to the energy deposited. Therefore, the
energy deposited by a shower at a given altitude is a function of
atmospheric properties (pressure, temperature, humidity, etc.) but it
is independent of the energy spectrum of the shower electrons.  The
validity of this assumption has been proven using both, theoretical and
experimental tools. In the following, the theoretical results are
summarized while experimental data are discussed later in
\sref{sec:Edep_exp}.

From a theoretical point of view several arguments have been used in
favor (with some caution) of an expected proportionality between
fluorescence intensity and deposited energy. Ave et al.
\Cite{5thFW_airfly_E} argue that since the total number of secondary
electrons produced by the passage of the primary electron in an air
volume is roughly proportional to the energy deposited, the
fluorescence light is also expected to be proportional to the energy
deposited. On the other hand, Arqueros et al.  \Cite{5thFW_Arqueros}
point out that the ratio of fluorescence emission and deposited energy
is strongly dependent on the spectrum of low-energy secondaries which
in principle varies with the primary energy and the distance from the
primary interaction.  Thus, this proportionality has to be
demonstrated with a detailed analysis.

The Monte Carlo algorithm in \Cite{5thFW_Arqueros} is used for a
detailed calculation of the energy dependence of the fluorescence
yield.  The results (see Fig.\,6 -- 9 in \Cite{5thFW_Arqueros}) can be
summarized as follows: the fluorescence yield decreases with primary
energy about 10\% in the range 1~keV -- 1~MeV and 4\% in the interval
1~MeV -- 20~GeV for the 337 nm band. For the 391 nm band the
corresponding decrease is about 6\% for the interval 1~keV -- 1~MeV
and 1\% for 1~MeV -- 20~GeV. A smooth increase of the fluorescence
yield with $P\times R$, smaller than 2\% in the range 15 --
1500~hPa$\times$cm, is found for energies larger than 1~MeV. At lower
energy and/or region size the fluorescence yield is clearly not
proportional  to the deposited energy.

In summary, the theoretical results in \Cite{5thFW_Arqueros} predict a
very small dependence of the fluorescence yield on electron energy
with no impact on the calibration of fluorescence telescopes.

\section{Compilation of data}
\Label{sec:data}

\subsection{Spectrum and pressure dependence}

As mentioned already, air-fluorescence emission  in the range 290
-- 430~nm is basically due to the 1N and 2P systems of N$_2$. The
spectrum consists of molecular bands with a degraded shape. The
wavelength and the intensity of these bands have been measured by many
authors. In particular, their spectral positions are accurately known
for many years, including the rotational structure. However,
relative intensities depend strongly on the gas features (i.e.\
pressure, gas composition, etc.) as given by \eref{eq:Rel_I} and
\eref{eq:1_P} -- \eref{eq:rel_v_bar}. Therefore, comparison between
different authors is not straightforward.

Although data on the relative intensities of the air-fluorescence
spectrum at several conditions have been published since long ago,
well established results are  not yet available. In the last years, the
air-fluorescence community has made a significant effort in achieving
accurate values of the relative intensities  and their pressure
dependence.

\begin{table*}[t]
\caption{Experimental results of $P^\prime$ at 293 K for the 2P and 1N
system of nitrogen in air. The first column shows the vibrational number
of the upper level of the transition with wavelength shown in the
second column. See text for a brief description of the main
experimental features. Uncertainties are not quoted.  More details can
be found in the original publications.}
\Label{P_prime}
\renewcommand{\thefootnote}{\thempfootnote}
\begin{minipage}{\linewidth}
\begin{center}
\begin{tabular}{ccccccccc}\hline
\multicolumn{2}{c}{2P system} & ~AIRFLY~\Cite{airfly_P}~ & \multicolumn{2}{c}{~Nagano et
al.~\Cite{nagano2004}\footnote{The
second column of Nagano et al.~values are weighted averages which are provided in their
publication.}~} &
~Bunner~\Cite{bunner1967}\footnote{
Weighted averages.
}~  & ~Pancheshnyi et al.~\Cite{pancheshnyi1998}\Cite{pancheshnyi2000}\footnote{
Inferred from quenching rate constant and lifetime measurements at 337~nm (2P $v = 0$), 316~nm (2P
$v = 1$), 314nm~(2P $v = 2$), 414~nm~(2P $v = 3$) and 391~nm~(1N $v = 0$).}
& ~AIRLIGHT~\Cite{5thFW_Waldenmaier}\footnote{
Inferred from quenching rate constants and lifetime measurements at 337~nm (2P $v = 0$), 316~nm (2P
$v = 1$), 391~nm (1N $v = 0$).}
~ & ~MACFLY~\Cite{MACFLY_thin}~ \\
v & $\lambda$ (nm) & \multicolumn{7}{c}{$P^\prime$ (hPa)} \\
\hline
\multirow{4}{*}{0} & 337.1 & 15.89 & 19.2 & \multirow{4}{*}{18.1} & \multirow{4}{*}{20.0}
& \multirow{4}{*}{13.10} & \multirow{4}{*}{15.0} & \multirow{4}{*}{25.8} \\
 & 357.7 & 15.39 & 18.1 & & & & & \\
 & 380.5 & 16.51 & 19.4 & & & & & \\
 & 405.0 & 17.80 & 12.3 & & & & & \\ \hline
\multirow{6}{*}{1} & 315.9 & 11.88 & 23 & \multirow{6}{*}{25.6} & \multirow{6}{*}{8.7}
& \multirow{6}{*}{11.20} & \multirow{6}{*}{15.0} & \multirow{6}{*}{17.1} \\
 & 333.9 & 15.50 & - & & & & & \\
 & 353.7 & 12.70 & 30.6 & & & & & \\
 & 375.6 & 12.82 & 34.1 & & & & & \\
 & 399.8 & 13.60 & 24.2 & & & & & \\
 & 427.0 & 6.38 & 72 & & & & & \\  \hline
\multirow{7}{*}{2} & 297.7 & 17.30 & - & \multirow{7}{*}{7.9} & \multirow{7}{*}{6.1}
& \multirow{7}{*}{9.10} & \multirow{7}{*}{-} & \multirow{7}{*}{11.4} \\
 & 313.6 & 12.27 & - & & & & & \\
 & 330.9 & 16.90 & 40.2 & & & & & \\
 & 350.0 & 15.20 & - & & & & & \\
 & 371.1 & 14.80 & - & & & & & \\
 & 394.3 & 13.70 & 24.2 & & & & & \\
 & 420.0 & 13.80 & 7.3 & & & & & \\   \hline
\multirow{3}{*}{3} & 296.2 & 18.50 & - & \multirow{3}{*}{-} & \multirow{3}{*}{3.3}
& \multirow{3}{*}{7.90} & \multirow{3}{*}{-} & \multirow{3}{*}{8.8} \\
 & 311.7 & 18.70 & - & & & & & \\
 & 328.5 & 20.70 & - & & & & & \\    \hline
\multirow{2}{*}{4} & 326.8 & 19.00 & - & \multirow{2}{*}{-} & \multirow{2}{*}{-}
& \multirow{2}{*}{-} & \multirow{2}{*}{-} & \multirow{2}{*}{-} \\
 & 385.8 & 19.00 & - & & & & & \\    \hline \\ \hline
\multicolumn{2}{c}{1N system} & & & & & & \\
\multirow{2}{*}{0} & 391.4 & 2.94 & 5.02 & 4.83 & 1.44 & 2.4 & 1.23 & 3.17 \\
 & 427.8 & 2.89 & - & - & - & - & - & - \\ \hline
1 & 388.5 & 3.9  & - & - & - & - & - & - \\ \hline
\end{tabular}
\end{center}
\end{minipage}
\end{table*}

Nagano et al. \Cite{nagano2004} have measured the relative intensities
of 15 bands in both pure nitrogen and dry air. As already mentioned,
in their experimental set-up excitation is carried out with electrons
from a 3.7 MBq $^{90}$Sr radioactive source and fluorescence is
spectroscopically resolved, using a set of interference filters.
Relative intensities are measured by electron-photon coincidences.
Values of the $P'_v$ parameters are determined from measurements of
the pressure dependence of fluorescence yield in the range 1 --
10$^3$~hPa using \eref{eq:Rel_I}.  Results on relative intensities and
characteristic pressures are given in Tabs.\,1 and 2 of
\Cite{nagano2004}. $P'$ values  are reviewed below in \tref{P_prime}
for comparison with other authors.  Notice that the determination of
the $P'_v$ values by Nagano et al.  does not take into account the
possible effect of secondary electrons escaping the field of view.
Since this effect is pressure dependent, it might give rise to
systematic uncertainties.

\begin{figure}[t]
\centering
\epsfig{width=0.49\textwidth, file=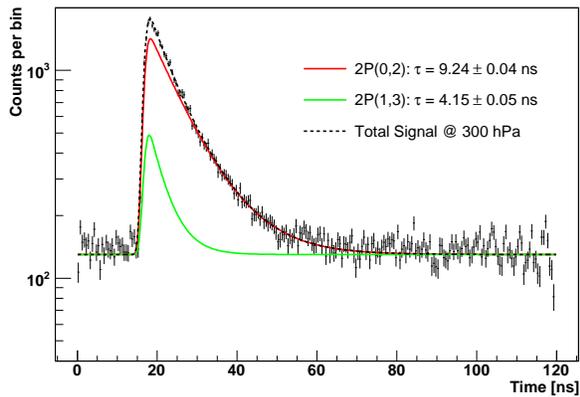}
\caption{Exponential time distribution between the electron
signals in the scintillator and the photon signals measured by the AIRLIGHT
experiment \Cite{5thFW_Waldenmaier}.}
\Label{5thFW_AirLight_fig04}
\end{figure}

The AIRLIGHT experiment \Cite{5thFW_Waldenmaier}, using a technique
similar to that of Nagano et al. \Cite{nagano2003}, have measured the
relative intensities of 8 nitrogen bands. However, here the $P'_v$
values are determined from the dependence of the fluorescence lifetime
on pressure. Delayed electron - photon coincidences are detected to
measure the exponential decay of the fluorescence emission, that is,
the effective lifetime (\fref{5thFW_AirLight_fig04}).

As already mentioned, the reciprocal lifetime increases linearly with
pressure (Stern-Volmer plot) and the corresponding slope provides the
$P'_v$ value as given by \eref{eq:tau_SV}.  Effective lifetimes at
high pressure are very low (in the range of few nanoseconds) and,
therefore, a high time resolution is necessary. This technique is free
from possible systematic uncertainties due to secondary electrons
escaping the field of view.

$P'_v$ measurements in dry air, pure nitrogen, and several mixtures
of nitrogen, oxygen, and water vapor allow Waldenmaier et al.
\Cite{5thFW_Waldenmaier} to determine quenching rate constants for the
various components separately, see \eref{eq:P_N} - \eref{eq:P_i}. The
results shown in Tab.~2 of \Cite{5thFW_Waldenmaier} can be used to
calculate the $P'_v$ values for any air-like mixture.  As an example,
the results for the 2P (v=0, 1) and 1N (v=0) transitions at 293 K are
inferred from the corresponding quenching rates and lifetimes and are
shown in \tref{P_prime} for comparison with other authors.

The AIRFLY Collaboration \Cite{5thFW_airfly_P}\Cite{airfly_P} has
achieved very accurate results on the air-fluorescence spectrum using
the electron beam of the Argonne Chemistry Van de Graaff facility. A
high resolution Oriel MS257TM spectrograph combined with a 1024
$\times$ 255 CCD pixel array (Andor DV420 BU2) allowed to record
high-resolution spectra of air fluorescence at 800 hPa and 293 K (see
\fref{airflyspec}). The spectral response was calibrated with an
uncertainty of 3\%. Using this technique, 34 bands of nitrogen in the
interval 280 -- 429~nm including a few weak lines of the Gaydon-Herman
system were identified.  The measured relative intensities are listed
in Tab.\,1 in \Cite{5thFW_airfly_P}. The experimental values for bands
with a common upper level are in good agreement with the theoretical
predictions given by the Einstein coefficients.

\begin{figure}[t]
\centering
\epsfig{width=0.49\textwidth, file=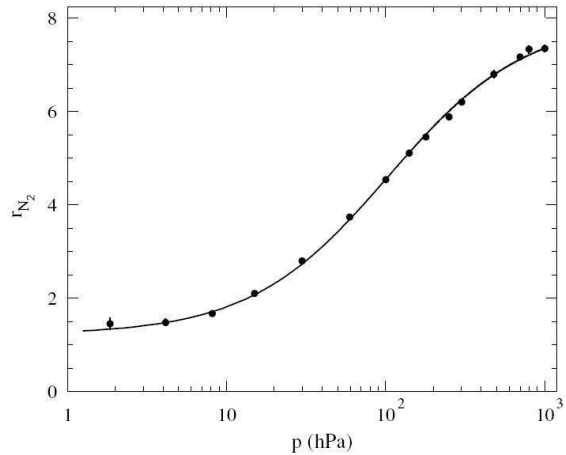}
\caption{The AIRFLY Collaboration determines the
P'$_{337}$ value for air from the ratio of nitrogen to air signals as a
function of pressure \Cite{5thFW_airfly_P}.}
\Label{fig3_AIRFLY_P}
\end{figure}

The measurements of the $P'_v$ values were performed at the Argonne
Wakefield Accelerator with 14 MeV beam energy, operated in pulsed
mode. The fluorescence intensity is recorded as function of pressure.
However, the procedure followed here is different from the one of
previous experiments e.g.\ \Cite{kakimoto1996}\Cite{nagano2004}.  AIRFLY
compares the pressure dependence of the 337 nm band for nitrogen and
dry air under the same experimental conditions (\fref{fig3_AIRFLY_P}).
Since the fraction of fluorescence losses due to secondary electrons
escaping the field of view
is basically the same for nitrogen and air, this comparison allows a
determination of $P'_{337}$ for quenching with N$_2$ and O$_2$ and,
hence, for air. An accurate result of $P'_{air}$=15.89$\pm$0.73 hPa is
reported in \Cite{5thFW_airfly_P}\Cite{airfly_P}.  The comparison
(with respect to the 337 nm line) of the pressure dependence for other
bands has provided a measure of $P'_v$ for other spectral components
of the air fluorescence light.

Following this technique, characteristic pressures of 25 bands of the
2P, 1N, and GH systems of nitrogen in air are reported in Tab.~2 of
\Cite{5thFW_airfly_P} with an uncertainty smaller than previous
experiments.  The results for the 2P and 1N systems have been included
in \tref{P_prime} for comparison with those of other authors.

The MACFLY experiment has measured the pressure dependence of the
total (unresolved) fluorescence intensity (290 -- 440 nm) for pure
nitrogen and air (Figs. 5 and 6 in \Cite{MACFLY_thin}).  The standard
theoretical formula ((4) in \Cite{MACFLY_thin}) is fitted to the
experimental pressure dependence,  using previously reported values of
the fluorescence yields at $P=0$ of pure nitrogen for 24
wavelengths, radiative lifetimes, and quenching rate constants at
$T=0$ for N$_2$ and O$_2$.  As a result, MACFLY reports dry-air $P'_v$
values for five upper levels, 2P ($v=0 - 3$) and 1N ($v=0$).  Notice
that these $P'_v$ results are not obtained from direct measurements
but are inferred as the most likely values, which are simultaneously
consistent with the experimental pressure dependence of the spectrally
unresolved fluorescence light yield and various experimental and
theoretical data on relative intensities and quenching from other
authors.  The results are also shown in \tref{P_prime} for comparison.

Finally, the FLASH Collaboration reports the measurement of the
air-fluorescence spectrum excited with electrons from the Final Focus
Test Beam at 28.5 GeV energy \Cite{flash_07}\Cite{5thFW_FLASH_thin}.
The spectrum was recorded using a spectrograph with about 1~nm
resolution. Fig.\,7 of \Cite{flash_07} shows an illustrative example
for air at 207 hPa.  Relative intensities are compared in Fig. 10 of
\Cite{flash_07} with other experimental values. Measurements of both
spectroscopically unresolved and resolved fluorescence intensity
versus pressure were carried out by the FLASH Collaboration (see
Tab.\,1 and Fig.\,8 in \Cite{flash_07}).

As summary, most of the available results on characteristic pressures
for the 2P and 1N nitrogen systems reported since the year 2000 are
compiled in \tref{P_prime}. Together with the measurements carried out
within the air-fluorescence community, the results of Pancheshnyi et
al.  \Cite{pancheshnyi2000} are shown for comparison. Most authors
provide values of the characteristic pressures for each $v$ value
using the experimental results of several molecular bands with the
same upper level $v$. Nagano et al. \Cite{nagano2004} and the AIRFLY
Collaboration \Cite{5thFW_airfly_P} report individual results for each
measured band. In both cases, $P'_v$ values reported for bands with
the same upper level $v$ are in agreement within the experimental
uncertainties, as well as with theoreticval expectations.

While \Cite{nagano2004} also provides averaged $P'_v$, AIRFLY
recommends using individual experimental results for each wavelength
as an empirical result independent of interpretations at molecular
level.

Uncertainties reported by the authors (not quoted in \tref{P_prime}) depend
strongly on the intensity of the molecular band.  As an example, the
uncertainty for the most intense band (337 nm) is below 1\% for the
AIRFLY measurement and below 3\% in Nagano et al.~\Cite{nagano2004}.
In regard with experiments reporting quenching rate constants for the
various upper levels, the uncertainty of AIRLIGHT
\Cite{5thFW_Waldenmaier} for 2P $v$=0 is around 5\% (assuming 4\%
uncertainty in the measurement of $K_Q$ for nitrogen) and more than
10\% for Pancheshnyi et al. \Cite{pancheshnyi2000}.

\subsection{Temperature dependence}
\Label{sec:temp_dep}

The fluorescence yield $Y_\lambda$ \eref{eq:FY} depends on temperature
through the $P^\prime_{v}$ parameter.  In general, the $P^\prime_{v}$
value for any mixture of gases like air can be written as introduced
in \eref{eq:1_P}, \eref{eq:P_i}, and \eref{eq:rel_v_bar}.  The well
known dependence on temperature results from the proportionality
between $P^\prime_{v}$ and $\sqrt{T}$. Assuming constant density, this
results in higher quenching rate constants for increasing temperature
because of the molecular Brownian motion.

From molecular physics, a second temperature dependence is expected.
The cross sections for collisional quenching $\sigma_{{\rm NN}}$ and
$\sigma_{{\rm NO}}$ depend on the velocity of molecules and therefore
their average value vary with temperature.  The expected behavior
follows a power law in temperature $\sigma\propto T^{\alpha}$, where
$\alpha$ is nearly constant in certain temperature intervals (see
\sref{sec:quenching}).  This $T$ dependence of the collisional cross
section had been neglected during the last decades as Bunner claimed
an evidence for only a weak temperature dependence \Cite{bunner1967}.
However, as reviewed in \Cite{histref}, considerable effects had been
found earlier. Recently, two groups started to measure the collisional
cross sections as a function of temperature.

The AIRFLY Collaboration measured at a Van de Graaff electron
accelerator with 3.0~MeV kinetic beam
energy~\Cite{5thFW_airfly_T}\Cite{AIRFLY_ICRC_Privitera}. The
fluorescence signal of dry air has been observed between 284 and
429~nm. A temperature scan with constant density was performed from
240~K to 310~K for the 2P bands 337~nm (0-0), 354~nm (1-2), 314~nm
(2-1), and the 1N band 391~nm (0-0).

\begin{table*}[t]
\caption{Measured temperature dependence parameters for a selected
         group of air-fluorescence bands as given by
     AIRFLY~\Cite{5thFW_airfly_T}\Cite{AIRFLY_ICRC_Privitera} and Fraga
         et al.~\Cite{5thFW_Fraga}. Additionally, the value extracted from
         Fig.~6 of \Cite{grun_schopper1954} for the entire wavelength range is
         given.  A measurement in the 1N system had been reported by
         Lillicrap~\Cite{lillicrap}.
         \Label{tab:alpha}}
\renewcommand{\thefootnote}{\thempfootnote}
\begin{minipage}{\linewidth}
\begin{center}
\begin{tabular}{ccccc}\hline
 & ~AIRFLY~\Cite{5thFW_airfly_T}\Cite{AIRFLY_ICRC_Privitera}~ & ~Fraga et
al.~\Cite{5thFW_Fraga}~ & ~Gr\"un et al.~\Cite{grun_schopper1954}~ &
~Lillicrap~\Cite{lillicrap}~ \\
 & in AIR & \multicolumn{3}{c}{in NITROGEN} \\
$\lambda$ (nm) & $\alpha_{v}$ &  $\alpha_{v}$ & $\alpha$~  &
$\alpha_{v}$\\
\hline
313.6 & -0.09 $\pm$ 0.10 & & & \\
337.1 & -0.36 $\pm$ 0.08 & -0.87 $\pm$ 0.15 & & \\
353.7 & -0.21 $\pm$ 0.09 & & & \\
391.4 & -0.80 $\pm$ 0.09 & & & -0.92\footnote{This value is valid in the temperature range
between 160 to 300~K. For lower temperature down to 78~K, $\alpha$ seems to change to
$-0.41$. }\\
$> 300$ & & & -0.79\footnote{This value has been obtained from a plot in
\Cite{grun_schopper1954}. The method is very imprecise and we do not know any
uncertainties for the data points in the plot. Thus, no uncertainties are provided for the
value.} \\ \hline
\end{tabular}
\end{center}
\end{minipage}
\end{table*}

In all cases, the fluorescence intensity is found to follow a power
law in the given temperature interval. For the 2P transitions, the
fluorescence signal decreases with temperature.  According to
\eref{eq:1_Ib}, this behavior implies that $\alpha > -1/2$ for the
corresponding upper level. However, the 1N fluorescence signal
increases with $T$ and, thus, $\alpha < -1/2$. The result of a fit to
the experimental data is shown in \tref{tab:alpha}.

Another experiment on the temperature dependence was performed by
Fraga et al., using pure nitrogen~\Cite{5thFW_Fraga}.  They use
$\alpha$-particles from a $^{241}$Am radioactive source.  After
several thorough tests of systematics, the temperature dependence of
the 337~nm band of nitrogen was  analyzed. During a temperature scan,
the density was kept constant as in the experiments from
AIRFLY~\Cite{5thFW_airfly_T}.  A fit to the measured data points
yields a value of $\alpha =  -0.87 \pm 0.15$ consistent with an
increasing quenching cross section with decreasing temperature.  This
result is also reviewed in \tref{tab:alpha} for comparison.  

The table also reports the measurements of older experiments in pure
nitrogen carried out by Gr\"un and Schopper~\Cite{grun_schopper1954}
as well as Lillicrap~\Cite{lillicrap}.
Gr\"un and Schopper measured the $T$ dependence for the integral
spectrum ($> 300$ nm). The $\alpha$-value reported in \tref{tab:alpha}
has been obtained from Fig.\,6 of \Cite{grun_schopper1954}. The result
of $\alpha = -0.79$ is consistent with that of Fraga et al., that is,
fluorescence increases with $T$ at a similar rate.  On the other hand,
Lillicrap reported measurements for the 1N (0-0) band (391~nm). In the
range 160 -- 300 K the temperature dependence indicates an 
$\alpha$-value of about $-0.92$, while at lower temperature (78 -- 160K)
$\alpha$ seems to increase up to about $-0.41$.  The AIRFLY
measurements for the 1N (0-0) band on air cannot be compared with that
of Lillicrap for pure nitrogen.  As pointed out in
\sref{sec:quenching}, the $\alpha$-parameter depends on the nature of
the partners and the type of interaction. In principle, its value for
N-O collisions is expected to be different than that for N-N
collisions. Since oxygen is a much more efficient quencher, the effect
of N-O collisions is expected to dominate in the temperature
dependence of the air-fluorescence light. Similarly, a comparison
between the results for the 337~nm band 2P(0-0) of Fraga et al.~and
AIRFLY is not possible.  

In \Cite{5thFW_Keilhauer}, the resulting fluorescence emission in
units of photons/m of a 0.85~MeV electron in the Earth's atmosphere is
shown for the AIRFLY data and Gr\"un and Schopper data. Because of an
additional $1/T$-dependence in the density-multiplication to calculate
the fluorescence emission in photons/m, the effect of
temperature-dependent collisional cross sections is strongest for the
Gr\"un and Schopper data. However, a caveat has to be applied to this
comparison: the $\alpha$-parameter from Gr\"un and Schopper has been
measured in pure nitrogen and is used there also for nitrogen-oxygen
quenching.

In summary, recent measurements confirm a temperature dependence of
collisional cross sections. Quenching cross sections decrease with
increasing temperature for both nitrogen-nitrogen and nitrogen-oxygen
collisions and for both molecular systems, 1N and 2P. Only the AIRFLY
Collaboration provides results for air. The comparison with
measurements on pure nitrogen is difficult since no information is
available on cross sections for N-O collisions.  Further
investigations are recommended because a relevance for cosmic-ray
measurements is indicated. Especially, measurements performed in one
set-up with both gases, nitrogen and air, as well as for
representative bands of the 2P and 1N system of nitrogen are highly
welcome to study the temperature-dependent collisional cross section
for nitrogen-nitrogen and nitrogen-oxygen quenching.

\subsection{Humidity dependence}
\Label{sec:hum_dep}

Water vapor is an always changing constituent of the Earth's
atmosphere. With respect to the fluorescence emission, the H$_2$O
molecules serve as an additional quenching partner for the excited
N$_2$ molecules. Hence, \eref{eq:1_P} has to be extended by a term
accounting for collisional quenching due to water vapor.

Measurements on water vapor quenching have been performed and reported
at the 5th Fluorescence Workshop by
AIRFLY~\Cite{5thFW_airfly_T}\Cite{AIRFLY_ICRC_Privitera}, Sakaki et
al.~\Cite{5thFW_sakaki}, and Waldenmaier et al.\
(AIRLIGHT)~\Cite{5thFW_Waldenmaier}\Cite{Waldenmaier2007}. Somewhat
earlier, measurements have been reported by Morozov et
al.~\Cite{morozov2005}, Pancheshnyi et
al.~\Cite{pancheshnyi1998}\Cite{pancheshnyi2000}, see e.g.\
\Cite{histref}, and the MACFLY Collaboration~\Cite{MACFLY_thin}.

\begin{table*}[t]
\caption{Values for $P^\prime_{{\rm H}_2{\rm O}}$ as measured by
several experiments. Since different experiments used different
techniques to obtain these values and some of them quote statistical
uncertainties only, here no uncertainties are reviewed.  Directly, the
values are incommensurable, thus, we ask the reader to refer to the
uncertainties in the original publications. All measurements were performed at
room temperature varying from about 288~K up to 300~K.
\Label{tab:pprime_hum}}
\renewcommand{\thefootnote}{\thempfootnote}
\begin{minipage}{\linewidth}
\begin{center}
\begin{tabular}{cccccc}\hline
 & ~AIRFLY~\Cite{5thFW_airfly_T}~ & ~AIRLIGHT~\Cite{5thFW_Waldenmaier}\footnote{Inferred
from measurements of quenching rate constants and lifetimes
.}~ & ~Morozov et al.~\Cite{morozov2005}\footnotemark[\value{mpfootnote}]~ & ~Pancheshnyi et
al.~\Cite{pancheshnyi1998}\Cite{pancheshnyi2000}\footnotemark[\value{mpfootnote}]~ &
~MACFLY~\Cite{MACFLY_thin}~ \\
 & \multicolumn{5}{c}{$P^\prime_{{\rm H}_2{\rm O}}$ (hPa)} \\
\hline
2P ($v$ = 0) & 1.28 & 1.92 & 1.31 & 2.47 & 2.94\\
2P ($v$ = 1) & 1.27 & 2.13 & 1.39 & 2.67 & 2.63\\
2P ($v$ = 2) & 1.21 & & & 2.59 & 2.55 \\
2P ($v$ = 3) & & & & 2.19 & 2.25 \\
1N ($v$ = 0) & 0.33 & 0.39 & & 0.76 & 0.76 \\ \hline
\end{tabular}
\end{center}
\end{minipage}
\end{table*}

The AIRFLY group measured the humidity dependence at a Van de Graaff
electron
accelerator~\Cite{5thFW_airfly_T}\Cite{AIRFLY_ICRC_Privitera}, as they
did it for the temperature, see \sref{sec:temp_dep}.  The fluorescence
chamber was filled with high purity dry air at atmospheric pressure.
Before the gas entered the chamber, it was flown through a bubbler
containing high purity water. The relative humidity in the chamber was
regulated from 0 to 100\% which is about 25~hPa partial pressure under
their experimental conditions.  In order to describe the water vapor
quenching, AIRFLY used the modified function for $1/P^\prime_{hum}$,
see \eref{eq:P_hum}.  They reported values for $P^\prime_{{\rm
H}_2{\rm O}}$ for the wavelengths 314~nm, 337~nm, 354~nm, and 391~nm,
see \tref{tab:pprime_hum}. This additional source of quenching has a
non-negligible effect, since the fluorescence yield decreases by about
20\% for a relative humidity of 100\% at atmospheric pressure and room
temperature.  

Sakaki et al.\ excited humid air with electrons from a $^{90}$Sr
source with 3.7~MBq, so that each run lasted about 1
week~\Cite{5thFW_sakaki}. The fluorescence yield was measured as
function of specific humidity for two bands of the 2P system (337 nm,
358 nm) and the most intense 1N band (391 nm).  A function
$\varepsilon(P)=C\times P /(1+P/P')$ is fitted to the data with $C$
and $P'$ as fitting parameters. This procedure allowed the authors to
find laws of photon yield versus altitude directly applicable to air
shower analyses.

Another experiment with $^{90}$Sr was performed by Waldenmaier et
al.~\Cite{5thFW_Waldenmaier}. Here the activity of the source was
37~MBq resulting in runs of about 30 hours. Only the last 20 hours
were used in the analysis to ensure stable conditions of the humidity
in the chamber. Overall, six runs at 15 -- 17$^\circ$C and 30~hPa of
pure nitrogen were obtained. Waldenmaier et al.~provide water vapor
quenching rate constants for 2P ($v=0,1$) and 1N ($v =0$), see
\tref{tab:pprime_hum}.

A comparison of the obtained values with Morozov et
al.~\Cite{morozov2005} and Pancheshnyi et
al.~\Cite{pancheshnyi1998}\Cite{pancheshnyi2000} shows that the
quenching rate constant from AIRLIGHT lies in between these two. The
values from AIRFLY agree quite well with those from Morozov et al.\
The relatively large values of Pancheshnyi et al.~might be due to
their method of mixing nitrogen with water vapor~\Cite{morozov2005}.
In their set-up oxygen and hydrogen were added to the nitrogen gas and
for the analysis it was assumed that the admixtures to nitrogen were
completely converted to water vapor in the cell. If not all hydrogen
and oxygen had been converted to water vapor, the actual water vapor
pressure could have been lower than expected, resulting in too high
$P^\prime_{{\rm H}_2{\rm O}}$ values. Also the $P^\prime_{{\rm
H}_2{\rm O}}$ values from MACFLY~\Cite{MACFLY_thin} are about twice
the values from AIRFLY resulting in less quenching and hence higher
fluorescence yield compared to AIRFLY. In the MACFLY experiment the
quenching rate constants have been calculated using a combination of
experimental data and molecular constants from results published
previously.

\subsection{Argon effect}
\Label{sec:argon}

Argon contributes to the Earth's atmosphere with 0.93\% per volume.
With respect to the nitrogen fluorescence emission, three possible
effects of argon have to be considered:  energy transfer from argon to
nitrogen,  direct fluorescence light emitted by argon, and collisional
quenching of excited nitrogen molecules with argon.

Argon can be excited from electrons by
\begin{equation}
{\rm e} + {\rm Ar} \rightarrow {\rm Ar}^\ast
\end{equation}
where the excitation cross section is largest for
Ar($^3P_2$)~\Cite{bennett_flint1978}. This excitation of argon is
followed by an efficient energy transfer from argon to nitrogen
\Cite{grun_schopper1954} via
\begin{equation}
{\rm Ar}^\ast + {\rm N}_2 \rightarrow {\rm Ar} + {\rm N}^\ast_2(C^3\Pi_u).
\end{equation}
The reached excited state of nitrogen is the known upper level of the
second positive system.

Under certain conditions, mixtures of argon with water vapor might
emit ultraviolet radiation at around 310~nm. The main contribution
comes from the transition OH $A^2\Sigma^+ \rightarrow X^2\Pi$
~\Cite{morozov2005a}.  A detailed study of this transition has shown
highest intensities for very low argon pressure and 0.06~Pa water
vapor.  Because of these special conditions, the fluorescence emission
of this transmission will be of no importance for the observation of
extensive air showers.

The increase of emission by energy transfer from argon to nitrogen
competes, however, with a higher quenching rate. The non-radiative
de-excitation of nitrogen is caused by additional collisions of
nitrogen with argon atoms in air.

Already very early Bunner stated the net effect of argon to be less
than 1\% contribution to the fluorescence light~\Cite{bunner1967}.
More recently, AIRFLY measured the effect of argon while comparing the
fluorescence yield from nitrogen-oxygen mixture with dry air. They
found that the effect of argon is completely negligible at atmospheric
pressure~\Cite{airfly_P}.

\subsection{Energy dependence}
\Label{sec:Edep_exp}

Theoretical results on the proportionality between fluorescence
intensity and deposited energy have been presented above
(\sref{sec:Edep}).  These predictions \Cite{5thFW_Arqueros} indicate
that the fluorescence yield is basically independent of the electron
energy for the typical experimental situations in this field.
Nevertheless, experimental tests of the proportionality between
deposited energy and fluorescence intensity are mandatory.

Several groups provided data on this topic. Two different experimental
methods have been used for this purpose. Firstly, in a thick-target
configuration the fluorescence intensity measured as function of the
shower depth is compared with the energy deposited by the shower,
either directly measured with an appropriate device or calculated with
a Monte Carlo code (e.g.\ EGS4, GEANT). Secondly, measurements of the
fluorescence yield in a thin-target experiment for various electron
energies provide the required data to test this proportionality.
Notice that in the latter technique it is necessary to measure the
fluorescence yield for well separated electron energies.

The FLASH collaboration measured the air fluorescence yield as a
function of the shower depth in a thick-target experiment for the
first time \Cite{flash_thick}. As mentioned above, this experiment
used 28.5~GeV electrons to induce an electromagnetic shower with a
composition similar to that generated by a 10$^{18}$ eV cosmic ray.
The authors found that the ratio of measured photomultiplier signals
to deposited energy is constant in the full shower depth range (2 --
14 radiation lengths) within 5\% uncertainty
(\fref{fig6_FLASH_thick}). The dependence of deposited energy on the
shower development stage, measured by an ion chamber, was found in
good agreement with a simulation carried out with EGS4.  Fig. 7 of
\Cite{5thFW_FLASH_thick} shows the relative value of the fluorescence
intensity as a function of depth for several band pass filters.  The
fluorescence intensity fits well the predictions of an empirical
deposition model, see (1) in \Cite{5thFW_FLASH_thick}.

\begin{figure}[t]
\centering
\epsfig{width=0.49\textwidth, file=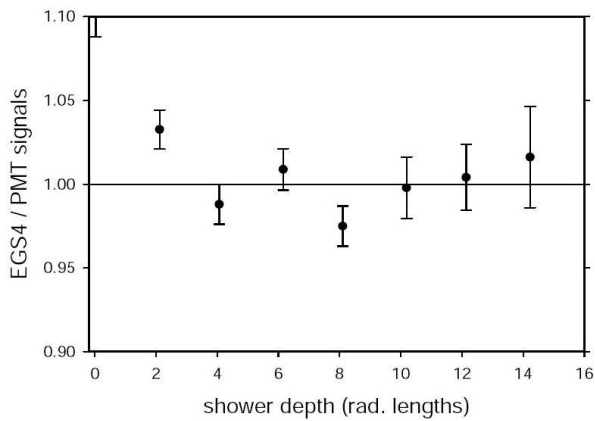}
\caption{Ratio of EGS4 predictions to weighted average of
         photomultiplier signals versus shower depth in the thick-target
         experiment of the FLASH Collaboration
\Cite{5thFW_FLASH_thick}.}
\Label{fig6_FLASH_thick}
\end{figure}

Several experiments have carried out direct measurements of the
fluorescence yield as a function of energy. The MACFLY experiment
measured the fluorescence yield at 1.5 MeV, 20 GeV, and 50 GeV using
the same collision chamber. MACFLY reports values of 17.0, 17.4, and
18.2 photons/MeV, respectively, with uncertainties of about 13\%. In
summary, the MACFLY thin-target experiment found that the fluorescence
yield is independent of energy in the interval 1.5 MeV -- 50 GeV.
Although no measurement was carried out for other energies inside this
large interval, no theoretical prediction suggests a lack of
proportionality inside this wide interval.

The MACFLY Collaboration carried out a thick-target experiment
\Cite{MACFLY_thick} which supports the above result. The MACFLY
thick-target experiment finds the fluorescence yield to be
proportional to the deposited energy for all stages of showers
initiated by high-energy electrons following a technique similar to
that of FLASH.

The AIRFLY Collaboration makes use of various accelerators to measure
the fluorescence yield in several energy intervals (see
\sref{sec:electron_sources} for more details). The obtained values for
the fluorescence yield are compared with the deposited energies
calculated with a full GEANT4 simulation of the experiment
\Cite{5thFW_airfly_E}. Since the inter-calibration between different
energy intervals has not yet been achieved, the result is only valid
inside energy intervals.

In the first place, the interval 0.5 -- 15 MeV is covered in two
subintervals 0.5 -- 3 MeV and 3 -- 15 MeV. In both cases the
fluorescence signal is proportional to deposited energy. In the second
one, the relativistic increase is clearly observed in the fluorescence
signal.  Since 3 MeV energy is included in both sub-intervals, the
proportionality test covers the whole interval 0.5 --  15 MeV.  A
deviation from perfect proportionality lower than 3\% is reported by
AIRFLY in this interval.  For the interval 50 -- 420 MeV, measurements
of the fluorescence yield turn out to be proportional to the deposited
energy also within 3\% uncertainty. Finally, the range 6 -- 30 keV is
studied, finding proportionality within 5\%.

In summary, AIRFLY shows that the fluorescence yield is independent of
electron energy within several energy intervals. Since these intervals
are very large, the AIRFLY data provide a good experimental test of
the proportionality assumption which should be completed in the future
by means of an absolute calibration in the whole energy range.

Other experiments reported proportionality between deposited
energy and fluorescence intensity although within much smaller energy
intervals. The AIRLIGHT experiment \Cite{5thFW_Waldenmaier} did not
find any noticeable dependence of the fluorescence yield on electron
energy in the range 0.25 -- 2.00 MeV. This test was carried out for
several molecular transitions in a pressure range of 50 -- 800 hPa. As
will be mentioned below, these authors carried out a detailed simulation
using GEANT4 to determine the energy deposited in the chamber.

Kakimoto et al.\ \Cite{kakimoto1996} compared the fluorescence
intensity measured in photons/m with the electron energy loss in air
for several energies in the range 1.4 MeV -- 1.0 GeV finding
proportionality.  A similar comparison of the photon yield (photons/m)
of Nagano et al. with the above results of Kakimoto et al. leads to
the same conclusion \Cite{nagano2003}. However, notice that in this
case the assumption has to be made that all energy lost by the
electrons is deposited within the field of view of the optical system.

\subsection{Absolute yield}
\Label{sec:Absolute}

The available results on absolute values of the air-fluorescence yield
are summarized in \tref{tab:absolute}.  As mentioned in
\sref{sec:yield}, different parameters can be used to account for the
fluorescence intensity emitted by an electron beam in air.  Some
experiments report results on the number of fluorescence photons per
electron and unit path length, i.e.\ the
$\varepsilon_{\lambda}$[m$^{-1}$] value, while other publications give
the result in number of photons per unit of deposited energy, i.e.\
the definition of fluorescence yield $Y_\lambda$ in this article.
Different experiments measure the fluorescence intensity in different
spectral intervals ranging from narrow band (e.g.\ 337 nm) to wide
spectral intervals (e.g.\ 290 -- 440 nm). In addition, very often
results are referred to different pressures and/or temperatures.

\begin{table*}[!t]
\caption{Experimental results on absolute values of the air-fluorescence
yield at room temperature and atmospheric pressure in the wavelength
interval shown in the third column. Measurements are carried out at
the energy given in the second column. Results are split depending on the
units used by the authors. For more details see text.}
\Label{tab:absolute}
\renewcommand{\thefootnote}{\thempfootnote}
\begin{minipage}{\linewidth}
\begin{center}
\begin{tabular}{lcccccccc}
\hline

Experiment  &  ~~~~~~~~ E  ~~~~~~~~ &  ~~~~ $\lambda$  ~~~~ & ~~~~ P ~~~~ &  ~~~~ T ~~~~
&\multicolumn{4}{c}{Fluorescence yield}\\
 & & & & &\multicolumn{2}{c}{~~wide spectrum~~}&\multicolumn{2}{c}{~~~~~~337 nm~~~~~~}\\

 &  [MeV]  &  [nm] & [hPa] & [K] & ~~~~[m$^{-1}$]~~ & ~~[MeV$^{-1}$]~~~~ & ~~~~[m$^{-1}$]~~ &
~~[MeV$^{-1}$]~~\\

\hline

AIRFLY \Cite{5thFW_airfly_A} &   350  &   337     &   993 &   291 &   ~   &   ~   &   ~   &
4.12\footnote{Preliminary, since the final absolute calibration is pending.}\\
FLASH \Cite{5thFW_FLASH_thin}    &2.85$\times$10$^4$& 300 -- 420 &   1013 &   304 &   ~   &
20.8 &   ~   &   ~\\
Lefeuvre et al. \Cite{Paris}  &   0.85  & 300 -- 430 &   1013 &   288 & 4.23  &  ~~   &   ~
&   ~\\
Nagano et al. \Cite{nagano2004}   &   0.85  & 300 -- 406 &   1013    &   293 & 3.81  &   ~
& 1.02  &   5.03 \\
~        &                  & 300 -- 430 &           &       & 4.05  &   ~   &       &   ~\\
MACFLY \Cite{MACFLY_thin}  &   1.5            & 290 -- 440 &   1013    &   296 & 3.14  &
17.6 &   ~   &   ~\\
~        &2.0$\times$10$^4$ &           &           &       & 4.22  &       &   ~   &   ~\\
~        &5.0$\times$10$^4$ &           &           &       & 4.44  &       &   ~   &   ~\\
AIRLIGHT \Cite{5thFW_Waldenmaier} &            0.25 -- 2.00  &  337      &   1013    &
293 &       &   ~   &       &  5.68 \\
Kakimoto et al. \Cite{kakimoto1996} & 1.4 -- 1000   & 300 -- 400  &    800   &   288    &       &
~   &       &  5.7 \\
\hline
\end{tabular}
\end{center}
\end{minipage}
\end{table*}

Kakimoto et al. \Cite{kakimoto1996} reported a value for the
fluorescence efficiency for the 337 nm band of $\Phi_{337}$ =
2.1$\times$10$^{-5}$ at 800 hPa and 288 K.  Therefore, the
fluorescence yield at these conditions equals 5.7 MeV$^{-1}$.
Kakimoto et al.\ used for the computation of $\Phi_{337}$ the energy
loss of the electron in the chamber as a measure of the deposited
energy.

Nagano et al. \Cite{nagano2004} provide the absolute number of photons
per meter at 293 K and an average energy of 0.85 MeV for individual
molecular bands, in particular the 337 nm one as well as the integral
value in two spectral intervals of interest (300 -- 406 nm and 300 --
430 nm). A fluorescence yield for the 337~nm band of 5.03 photons/MeV
at 293 K and 1013 hPa is easily inferred from the $\Phi^0_{337}$ and
$P'_{337}$ values reported in this work. In principle, the possible
effect of secondaries escaping the field of view (see
\sref{sec:abs_calibration}) is not treated in this work. The quoted
uncertainty is 13\%.

At the same average energy Lefeuvre et al. \Cite{Paris} report a value
of 4.23 photons/m in the range 300 -- 430 nm at 288 K and 1013 hPa.
These measurements lack from a systematic study of the pressure
dependence of the fluorescence yield and, therefore, they cannot
provide $P'$ values. Although the wavelength spectrum was registered
using a monochromator, relative intensities were not measured and,
consequently, the absolute value of the fluorescence yield for the 337
nm band was not reported.  According to their calculations, a very
small correction for the effect of lost secondary electrons has to be
applied. The authors claim to have achieved an extremely high accuracy
(5.0\% uncertainty).

The AIRLIGHT experiment \Cite{5thFW_Waldenmaier} provides results on
the quenching rate constants and lifetimes for several molecular
bands. The energy deposited in the chamber was calculated using GEANT4
showing that a non-negligible correction is necessary, in particular,
at high energy and high pressure. The authors give results on the
fluorescence yield at zero pressure with a systematic uncertainty of
about 15\% which, combined with quenching rate constants and
lifetimes, allows to determine the absolute value of the
air-fluorescence yield for any pressure and temperature for several
molecular bands. As an example, the predicted value at 293 K and
atmospheric pressure for the 337 nm band is shown in
\tref{tab:absolute}.

Three experiments have carried out absolute measurements of the
air-fluorescence yield using high-energy electrons.  The FLASH
Collaboration, working with 28.5 GeV electrons, reports absolute
values in the spectral range 300 -- 420 nm at several pressures in the
interval 67 -- 1013 hPa and 304 K temperature. The value at
atmospheric pressure is 20.8 photons/MeV. The authors performed an
absolute calibration by comparison with the Rayleigh scattering from a
nitrogen laser. An uncertainty of this absolute value below 8\% is
reported.

The MACFLY Collaboration measures the absolute fluorescence yield in
the spectral range 290 -- 440 nm at 296 K. The authors report results
in photons/m at three electron energies, 1.5 MeV, 20~GeV, and 50 GeV.
A comparison of these numbers with the energy deposited in the
chamber according to a GEANT4 simulation leads to an averaged
fluorescence yield value of 17.6 photons/MeV for the above mentioned
spectral range and temperature. According to the authors, the
uncertainty of this result is about 15\%.

Finally, the AIRFLY Collaboration carried out systematic studies
on the dependence of the fluorescence yield on pressure and electron
energy. As mentioned above, this collaboration has developed a novel
method for the absolute calibration of the experimental system based
on the comparison with \Cerenkov radiation generated by the
electron beam. The AIRFLY Collaboration carried out a detailed GEANT4
simulation to determine accurately the energy deposited inside the
field of view of the optical system.  A careful analysis of the
various contributions to the systematic uncertainties leads them to
conclude conservatively that reducing the uncertainty below the 10\%
level is achievable. Using an electron beam of 350 MeV, a preliminary
result of 4.12 photons/MeV for the 337 nm band at 291 K is given.
Further measurements and checks will lead in a near future to a final
result for the absolute value.

A comparison of experimental results when expressed in different
units is not straightforward. Assuming a fixed wavelength interval,
pressure and temperature, results in photons/m could be translated to
photons/MeV as far as the deposited energy (within the field of view
of the optical system) per unit path length and electron is known.

Results on this parameter have been calculated using EGS4 or GEANT4
for several experimental configurations
\Cite{MACFLY_thin}\Cite{5thFW_FLASH_thin}. In \Cite{5thFW_Arqueros}, a
simulation at microscopic level gives results in good agreement with
those mentioned above.  The results of \Cite{5thFW_Arqueros} indicate
that deposited energy per unit column density is only smoothly
dependent on the product of pressure and radial size of the
observation volume.

For the comparison of different measurements in the same units,
assuming similar geometry, but in different spectral ranges, the
relative intensities and the characteristic pressures are needed.

In \Cite{5thFW_Arqueros}, a procedure for the comparison of absolute
values expressed in different units and for different spectral ranges
based on the above arguments is shown. Since data on relative
intensities, $P'_v$ values, and deposited energy per meter in a
particular configuration have uncertainties, this procedure does not
allow to compare high-accuracy measurements, nevertheless it can
be useful for a comparison of experimental data at the level of about
15\% uncertainty.

\section{The fluorescence light yield in the atmosphere}
\Label{sec:altitude}

For cosmic-ray experiments, the measurement of the nitrogen
fluorescence emission is the most direct method to detect the
longitudinal profile of extensive air showers. For the event
reconstruction procedures of these air shower experiments, the
knowledge of the fluorescence yield and its dependence on atmospheric
conditions are crucial parameters. The principle of air shower
detection with fluorescence light has been discussed above
(\sref{sec:principle}). The most relevant altitude range for ultra
high-energy cosmic rays is between ground level and about 13~km above
sea level (a.s.l.). The shower maximum is reached between 2 and
8~km~a.s.l. for a shower with 10$^{19}$~eV, depending on type and
inclination angle of the primary particle.  The field of view of the
fluorescence telescopes of air shower experiments covers this range.
For example, the Auger telescopes oversee the sky between 0.7~km and
12.5~km above the altitude of the Pierre Auger Observatory,
1.4~km~a.s.l., at a distance of 20~km.

Up to at least 80~km~a.s.l., it is safe to assume a constant
composition of the Earth's atmosphere which is mainly 78.08\% N$_2$,
20.95\% O$_2$, and 0.93\% Ar per volume. All three constituent parts
influence the emission of fluorescence light, however, with strongly
differing importance. As already discussed throughout this summary
article, emission from nitrogen is the dominant light.  The
contribution of argon has been discussed in \sref{sec:argon}. The
UV-fluorescence light emission from O$_2$ is
negligible~\Cite{nicolls_etal1959}. The contribution between 300 and
400~nm stems from O$^+_2A^2\Pi_u$ - $X^2\Pi_g$ transitions.  However,
their intensities are negligible as compared with those of nitrogen.
The emission of atomic oxygen has wavelengths larger than 395~nm, up
to 845~nm~\Cite{nicolls_etal1959}, but with no relevance for air
shower experiments.

To compare the fluorescence light profiles in the Earth's atmosphere,
in former discussion often an electron energy of 0.85~MeV has been
chosen.  These electrons lose their energy mainly by ionization and
the energy deposit is $dE/dX$ =
0.1677~MeV/kg~m$^{-2}$~\Cite{nagano2003}. To obtain a light profile
induced by an extensive air shower, the longitudinal profile of the
locally deposited energy $E_{dep}$ per g~cm$^{-2}$ has to be known. A
resultant number of fluorescence photons $\varepsilon_\lambda$ with
wavelength $\lambda$ per unit path length is then given as
\begin{equation}
\varepsilon_\lambda = Y_\lambda \cdot \frac{\lambda}{hc}\cdot E_{dep} \cdot \rho_{air}.
\end{equation}
The air density $\rho_{air}$ is that at the position of the energy
deposit, thus, where the photons are emitted. $Y_\lambda$ is the
fluorescence yield including all temperature, pressure, and humidity
dependences as discussed in \srefs{sec:temp_dep}, \Ref{sec:Edep_exp},
and \Ref{sec:hum_dep}, respectively.  With regard to extensive air
showers, the transition from the fluorescence yield $Y_\lambda$ in
photons emitted per deposited energy to the number of fluorescence
photons $\varepsilon_\lambda$ per unit path length is straightforward
because all secondary electrons involved in the fluorescence
process are in the observed volume. No boundary effects as for
laboratory measurements have to be considered as discussed in
\sref{sec:fl_methods}.

\begin{figure}[t]
\begin{center}
\includegraphics*[width=0.47\textwidth,angle=0,clip]{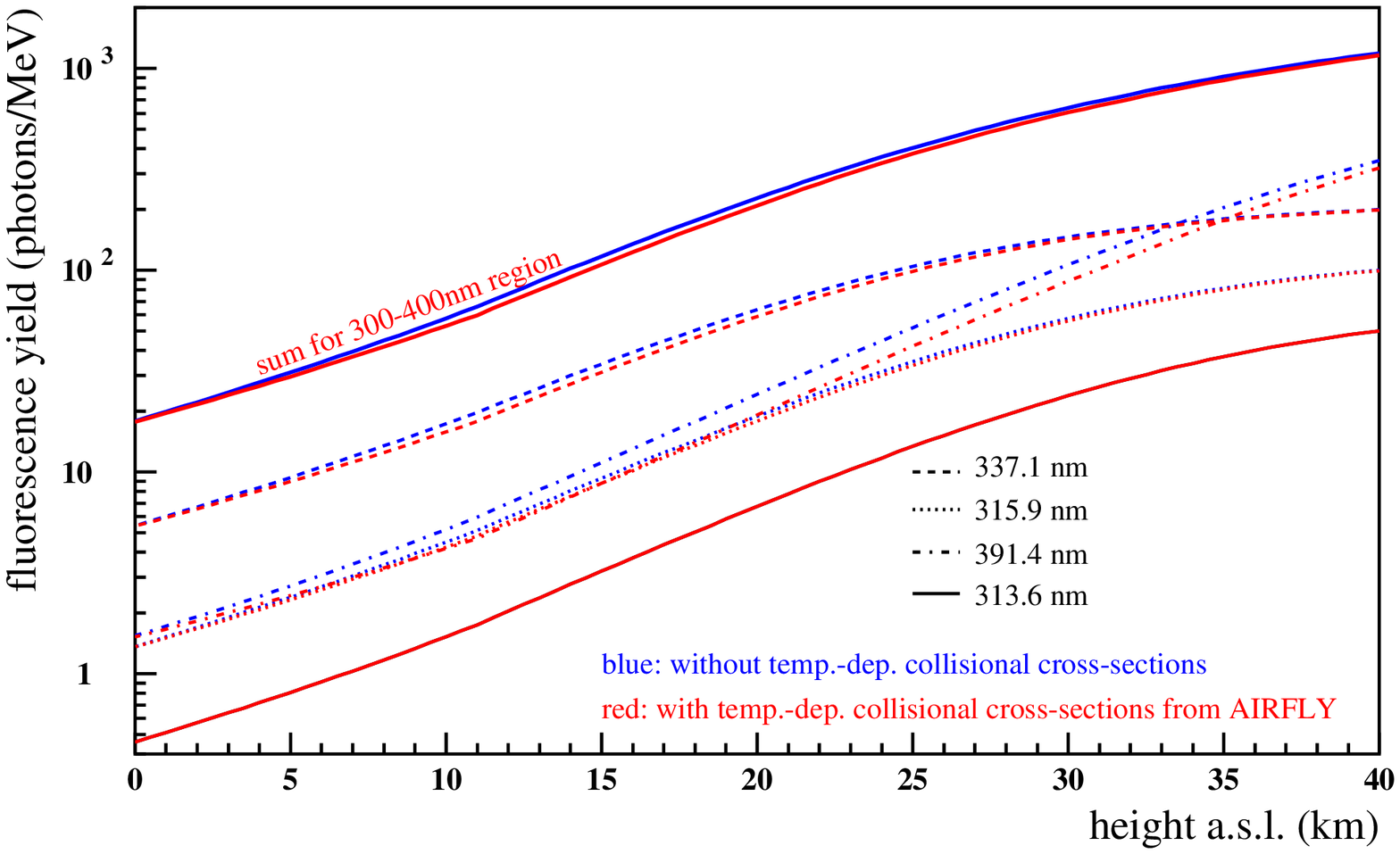}
\caption{\Label {fig:yield_comparison_tempdep}
Fluorescence yield profile for different wavelengths in the US
Standard Atmosphere. The blue curves represent the fluorescence yield
accounting for the $\sqrt{T}$-dependence only, compare to Fig.~3 in
\Cite{keilhauer2006}. The red curves account for the entire
temperature dependence, thus including the temperature-dependent
collisional cross sections~\Cite{5thFW_Keilhauer}, with
parametrization given from AIRFLY~\Cite{AIRFLY_ICRC_Privitera}. }
\end{center}
\end{figure}

Before this workshop, only the $\sqrt{T}$-dependence and the pressure
dependence have been taken into account. In the Earth's atmosphere as
described in the US Standard Atmosphere~\Cite{US-StdA}, the
temperature decreases up to 11~km~a.s.l.~with a continuous lapse rate
of 6.5~K/km. For higher altitudes, until 20~km~a.s.l., the temperature
remains constant. Above, the temperature increases again until the
stratopause at around 50~km~a.s.l. This temperature profile, together
with the pressure profile of the atmosphere, affects the fluorescence
yield in the atmosphere differently for each band system due to
different deactivation constants. In
\fref{fig:yield_comparison_tempdep}, the blue curves represent the
fluorescence yield accounting for the
$\sqrt{T}$-dependence~\Cite{keilhauer2006}.

\begin{figure}[t]
\begin{center}
\includegraphics*[width=0.47\textwidth,angle=0,clip]{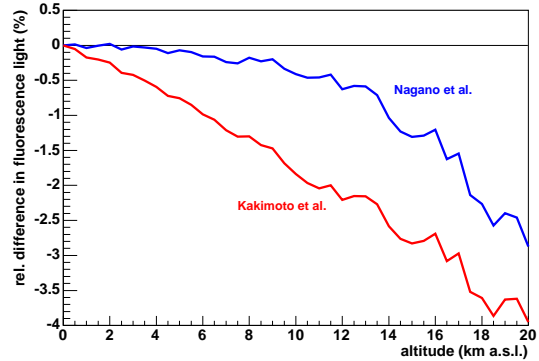}
\caption{\Label{fig:yield_hoehe_N_KUl_Kak_relVgl}
Relative difference for the fluorescence light profiles in the US
Standard Atmosphere with a 0.85~MeV electron as incident particle of
the two parametrizations, \eref{eq:alt_param_Nagano} and
(\Ref{eq:FY_N}) compared with the detailed calculation of
\Cite{keilhauer2006}.}
\end{center}
\end{figure}

\begin{table}[t]
\caption{$A$ and $B$ values in 10 bands between 300 and 400~nm as
given by Nagano et al.~\Cite{nagano2004}.\Label{tab:nagano}}
\begin{minipage}{\linewidth}
\begin{center}
\begin{tabular}{ccc}\hline
Main $\lambda$ (nm) & ~$A$ (m$^2$kg$^{-1}$)~ & ~$B$ (m$^3$kg$^{-1}$K$^{-1/2}$)~ \\
\hline
316 & 20.5 $\pm$ 1.3 &  2.14 $\pm$ 0.18 \\
329 & 3.91 $\pm$ 0.35 & 1.22 $\pm$ 0.14\\
337 & 45.6 $\pm$ 1.2 & 2.56 $\pm$ 0.10\\
354 & 3.68 $\pm$ 0.39 & 1.60 $\pm$ 0.21\\
358 & 37.8 $\pm$ 2.3 &  2.72 $\pm$ 0.22 \\
376 & 6.07 $\pm$ 0.57 & 1.44 $\pm$ 0.17\\
381 & 12.7 $\pm$ 1.4 & 2.53 $\pm$ 0.35\\
391 & 50.8 $\pm$ 2.1 & 9.80 $\pm$ 0.51\\
394 & 2.25 $\pm$ 0.78 & 2.03 $\pm$ 0.79\\
400 & 4.58 $\pm$ 0.44 & 2.03 $\pm$ 0.23\\
\hline
\end{tabular}
\end{center}
\end{minipage}
\end{table}

A detailed calculation of the fluorescence emission including
atmospheric effects with constant collisional cross sections
$\sigma_{{\rm N}x,v}$ has been compared in Keilhauer et
al.~\Cite{keilhauer2006} with parametrizations of the altitude
dependence. Nagano et al.~provided parameters $A_\lambda$ and
$B_\lambda$ for all 10 wavelengths they measured between 300 and
400~nm, see \tref{tab:nagano}, describing the fluorescence light
emission per unit path length as~\Cite{nagano2004}
\begin{equation}
\Label{eq:alt_param_Nagano}
\varepsilon_\lambda = E_{dep}\cdot \biggl(\frac{A_\lambda \rho}{1+\rho B_\lambda
\sqrt{T}}\biggr).
\end{equation}
Kakimoto et al.~reported only one set of parameters $A_{1,2}$ and $B_{1,2}$,
see Tab.~12 in \Cite{histref}, in order to describe the entire wavelength
range with \eref{eq:FY_N}.  Comparing only the altitude dependence in the
US Standard Atmosphere~\Cite{US-StdA}, assuming an equal absolute number of
fluorescence photons at ground, shows a quite good agreement, see
\fref{fig:yield_hoehe_N_KUl_Kak_relVgl}.

A further analysis of the altitude dependence has been performed for
the atmospheric conditions at the site of the Pierre Auger Observatory
\Cite{keilhauer2006}. Four seasonal atmospheric models have been
developed. The fluorescence light induced by a 0.85~MeV electron,
applying the well-known dependences, has been compared to that
expected in the US Standard Atmosphere, see Fig.~5
in~\Cite{5thFW_Keilhauer}\footnote{In the nomenclature of this summary
article, the fluorescence yield shown there corresponds to
$\varepsilon_\lambda$.}.  The differences for the Argentine seasons
compared with the US Standard Atmosphere are well below
$\pm$5\%~\Cite{keilhauer2006}.

\begin{figure}[t]
\begin{center}
\includegraphics*[width=0.47\textwidth,angle=0,clip]{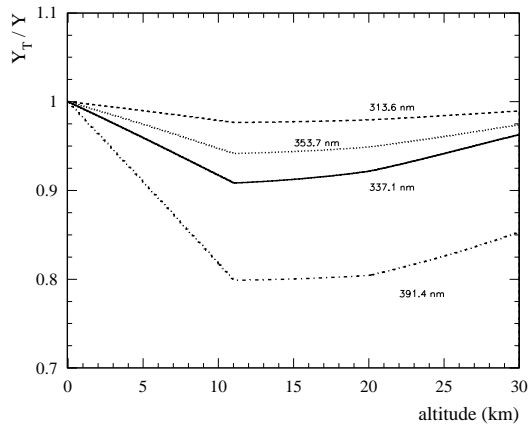}
\caption{\Label {fig:airfly_alpha} Ratio of fluorescence yield with
measured $\alpha_\lambda$ to the one with $\alpha_\lambda$ = 0: dashed
line 313.6~nm; full line 337.1~nm; dotted line 353.7~nm; dashed-dotted
line 391.4~nm~\Cite{5thFW_airfly_T}.}
\end{center}
\end{figure}

Introducing also the effect of temperature-dependent collisional cross
sections, see \sref{sec:temp_dep}, the reduction of the fluorescence
yield is significant. The red curves in
\fref{fig:yield_comparison_tempdep} are the fluorescence yield
accounting for both temperature effects with data from AIRFLY. The
relative differences of the yield with both effects compared to the
fluorescence yield just with the $\sqrt{T}$-dependence can be seen in
\fref{fig:airfly_alpha}.
The overall reduction for the wavelength range between 300 and 400~nm
is dominated by the reduction of the 337~nm band up to 18~km~a.s.l.,
so that these two curves would lie on top of each other. For higher
altitudes, the contributions of other bands become more important and
the overall reduction increases compared to that of the 337~nm band.

Applying this dependence to the calculation of the light emission
profiles of extensive air showers, the expected shower light profile
of an Fe-induced cascade with 10$^{19}$~eV is reduced by 2.7\% up to
7.5\% depending on inclination of the shower and the atmospheric
model~\Cite{5thFW_Keilhauer}. The values given here refer to Argentine
seasonal atmospheric models as used for the Pierre Auger
Observatory~\Cite{keilhauer2004}.

\begin{figure}[t]
\begin{center}
\includegraphics*[width=0.47\textwidth,angle=0,clip]{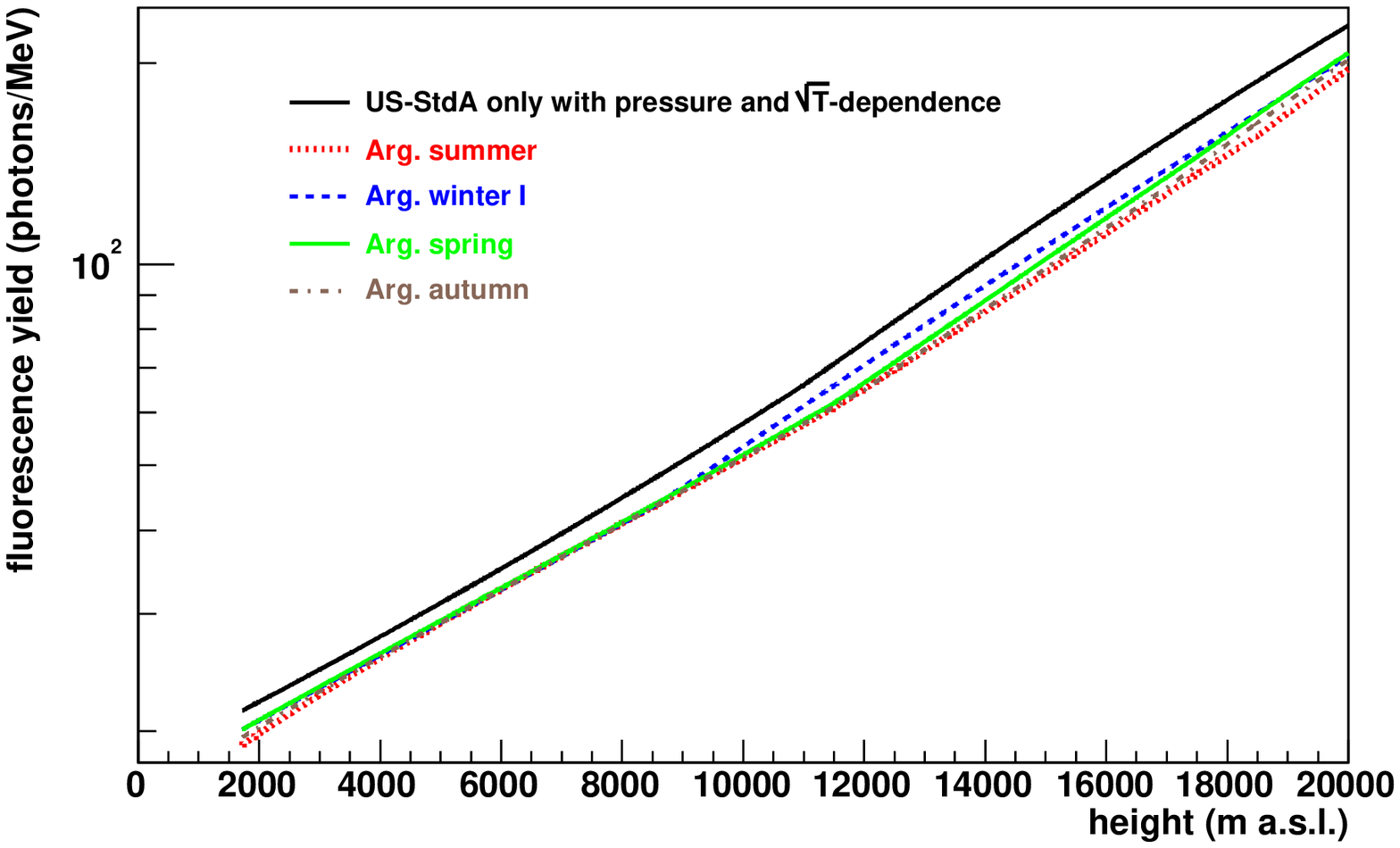}
\caption{\Label{fig:fl_yield_altdep_logplot} Fluorescence yield $Y_\lambda$ in the US Standard
Atmosphere only with pressure and $\sqrt{T}$-dependence together with the fluorescence
yield as expected in Argentine atmospheres accounting for all newly known dependences. The
temperature-dependent collisional cross sections are taken from
AIRFLY~\Cite{5thFW_airfly_T}\Cite{AIRFLY_ICRC_Privitera} and the collisional cross sections due
to water vapor are measured by Waldenmaier et
al.~\Cite{5thFW_Waldenmaier}\Cite{Waldenmaier2007}.}
\end{center}
\end{figure}

\begin{figure}[t]
\begin{center}
\includegraphics*[width=0.47\textwidth,angle=0,clip]{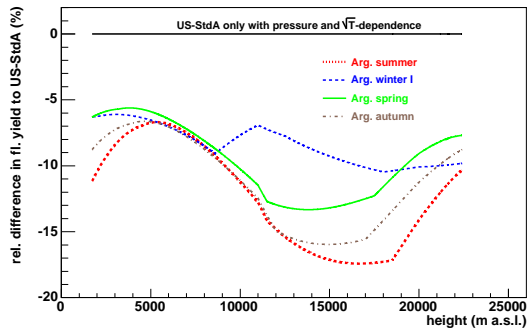}
\caption{\Label{fig:dpyield_altdep}
Relative difference of the fluorescence yield $Y_\lambda$ in Argentine
atmospheres with the latest results of pressure, temperature, and
humidity dependences to that in the US Standard Atmosphere only with
pressure and $\sqrt{T}$-dependence.}
\end{center}
\end{figure}

These seasonal atmospheric models have been used to study further the
altitude dependence of the fluorescence yield. In order to include the
humidity dependence, seasonal average profiles of the relative
humidity at the site of the Pierre Auger Observatory have been
obtained by meteorological radio soundings during
night-time~\Cite{5thFW_Keilhauer}.  As a result of the newly measured
temperature and humidity dependences, \sref{sec:temp_dep} and
\Ref{sec:hum_dep}, combined with the well-known
dependences on temperature and pressure, also \sref{sec:temp_dep} and
\Ref{sec:Edep_exp}, the fluorescence yield $Y_\lambda$ changes
significantly in the Earth's atmosphere.
\fref{fig:fl_yield_altdep_logplot} displays $Y_\lambda$ in the US
Standard Atmosphere only with pressure and $\sqrt{T}$-dependence
together with the fluorescence yield as expected in Argentine
atmospheres accounting for all newly known dependences.
The relative difference of the fluorescence yield with the latest
results to that in the US Standard Atmosphere with the well-known
dependences is of the order of 10\%, as can be seen in
\fref{fig:dpyield_altdep}.

\begin{figure}[t]
\begin{center}
\includegraphics*[width=0.47\textwidth,angle=0,clip]{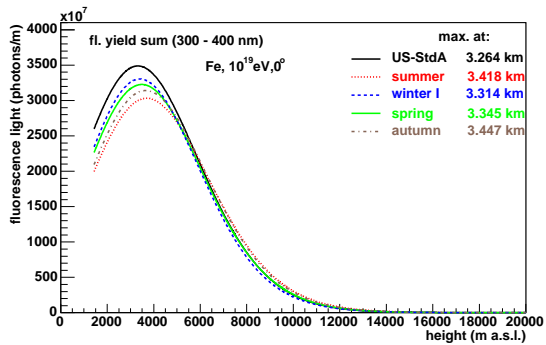}
\caption{\Label{fig:fl_yield_E19_0deg_wVapdep_USodep}
Fluorescence light profiles for average iron-induced air showers with
10$^{19}$~eV with vertical incidence~\Cite{5thFW_Keilhauer}. For the
profiles in Argentine atmospheres, all newly measured altitude
dependences are taken into account. For the US Standard Atmosphere
only the pressure and $\sqrt{T}$-dependence have been considered.}
\end{center}
\end{figure}

\begin{figure}[t]
\begin{center}
\includegraphics*[width=0.47\textwidth,angle=0,clip]{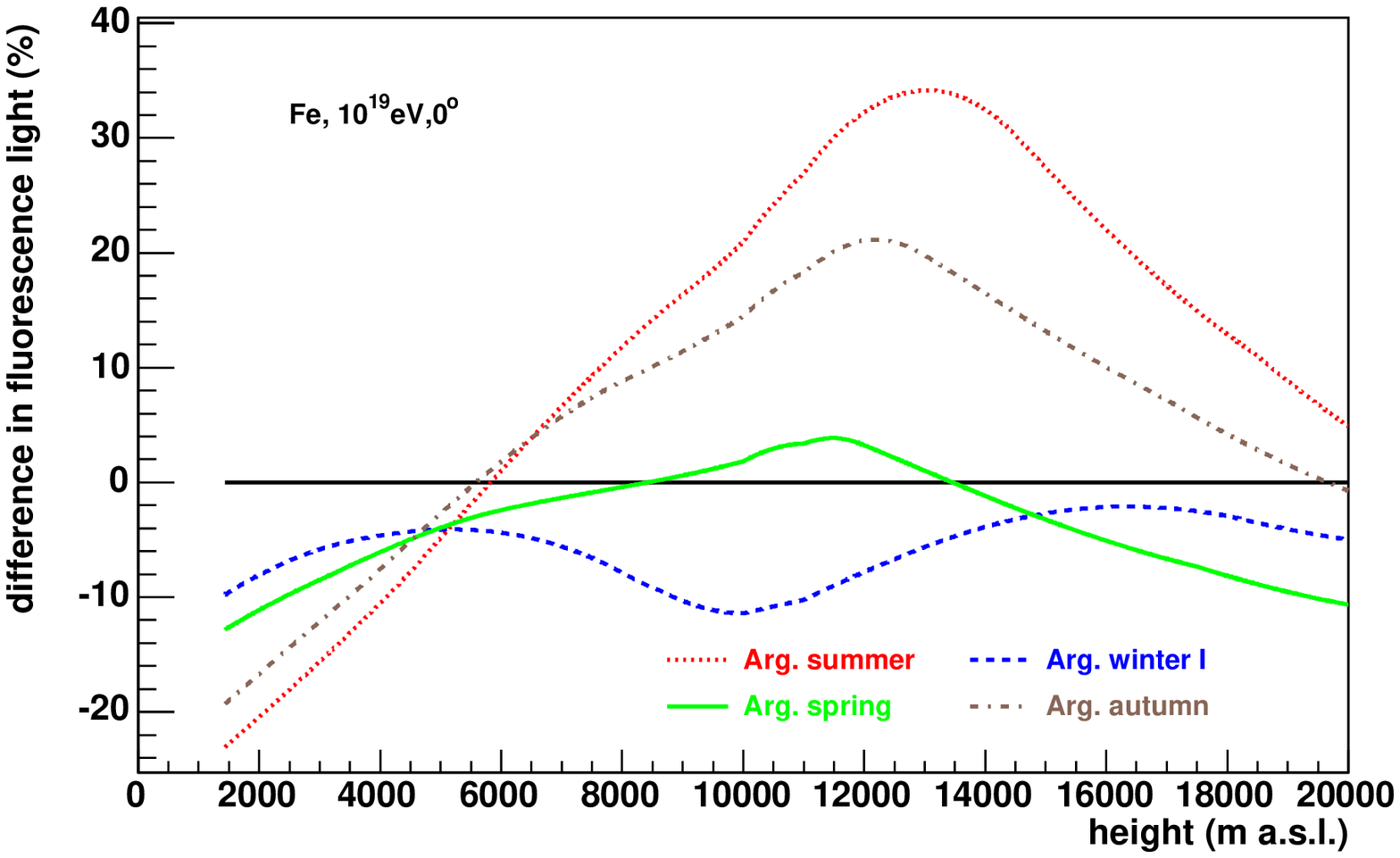}
\caption{\Label{fig:dpyield_E19_0deg_wVapdep_USodep} Difference of the fluorescence light
profiles in Argentine atmospheres to that in the US Standard Atmosphere as shown in
\fref{fig:fl_yield_E19_0deg_wVapdep_USodep}~\Cite{5thFW_Keilhauer}.}
\end{center}
\end{figure}

To estimate the importance of the newly measured altitude
dependences on reconstructing shower profiles, average iron-induced
air showers with $E_0 = 10^{19}$~eV have been simulated with
CORSIKA~\Cite{CORSIKA} in \Cite{5thFW_Keilhauer}. The fluorescence
light profile for the case of vertical incidence can be seen in
\fref{fig:fl_yield_E19_0deg_wVapdep_USodep} and the corresponding
difference in \fref{fig:dpyield_E19_0deg_wVapdep_USodep}.
The expected light of the air shower is reduced by 11.1\% during
summer, 8.9\% during autumn, 7.3\% during spring, and 6.8\% during
winter~\Cite{5thFW_Keilhauer}. For the same shower with 60$^\circ$
inclination, the expected light is reduced by 8.4\% during winter,
8.1\% during spring and autumn, and 8.0\% during summer. The same
calculations have been performed for proton-induced air showers. The
reduction of the expected light is increased by about 0.5\% compared
to the numbers of the iron-induced cascade \Cite{5thFW_Keilhauer}.
Hence, accounting for the newly measured altitude dependences in the
reconstruction of the primary energy of the shower, the primary energy
will be increased by this amount as compared with the former model
calculations.

\section{Summary and Outlook}
\Label{sec:outlook}

The 5th Fluorescence Workshop has brought together experimental and
theoretical expertise from around the world to discuss the status of
the determination of the fluorescence light yield of electrons,
important for air shower observations. It can be realized with
pleasure that over the last years significant progress has been
achieved in both, experimental and theoretical work.

Convergence has been achieved in several aspects between the different
groups.  Most important for air shower observations is the
fluorescence light yield of nitrogen, since the contributions of direct
excitation of oxygen and argon seem to be negligible.
Hence, investigations focus on the fluorescence light emission of
nitrogen. The position of the molecular bands in the fluorescence
light spectrum is well known. Accurate measurements of the relative
intensities at typical laboratory conditions have been presented. Also
high precision values of the characteristic pressures have been
reported, although full agreement is still missing. High-precision
$P'$ values allow accurate predictions on the relative fluorescence
intensity at any atmospheric condition of interest for air shower
reconstruction. The proportionality between fluorescence intensity and
deposited energy seems to be proved from both experimental and
theoretical sides. Recent investigations reveal that the dependence of
the light yield on temperature and humidity can not be neglected. In
realistic atmospheres effects up to $10\%$ can be expected. Further
measurements are necessary to clarify the situation.

A precise determination of the absolute light yield is extremely
difficult and still needs more experimental data.  For a comparison of
the values obtained by different groups, it would be useful if the same
concept is used. The participants agreed to specify the fluorescence
light yield in photons per energy deposition (photons/MeV). This
quantity  is most useful to convert the observed light yield into
a calorimetric measurement of the energy of an air shower. The usage
of this unit is  strongly encouraged.

The open questions are presently addressed by several groups. The
workshop participants expressed their hope to achieve convergence on the
open issues soon ($1-2$ years).

\section*{Acknowledgements}

The authors acknowledge the support of the Spanish Ministry of Science
and Education MEC (FPA2006-12184-C02-01), Comunidad de Madrid (Ref.:
910600) and CONSOLIDER program, and of the German Research Foundation
(DFG) (KE 1151/1-1 and KE 1151/1-2). Further support has been granted
by the Radboud Universiteit Nijmegen as well as by the Universit\"at
Karlsruhe (TH) and the Forschungszentrum Karlsruhe GmbH which are
currently merging their activities in the Karlsruhe Institute of
Technology (KIT). The authors would like to thank C.~Escobar,
C.~Field, M.~Fraga, K.~Martens, M.~Nagano, J.~Ridky, J.~Rosado, and
A.~Ulrich for valuable comments on the manuscript.

%\bibliographystyle{elsart-num}
%\bibliography{biblio}

\end{document}